\documentclass[10pt,journal,compsoc]{IEEEtran}

\usepackage{bbding}
\usepackage{pifont}
\usepackage{amssymb}
\usepackage{multirow}
\usepackage{booktabs}
\usepackage{listings}
\usepackage{multirow}
\usepackage{pifont}
\usepackage{caption}
\usepackage{hyperref}
\usepackage{tabularx}
\usepackage{xr}
\usepackage{balance}
\usepackage{color}
\usepackage{soul}
\usepackage{bm}
\usepackage{amsmath}
\usepackage{xfrac}
\usepackage{multirow}
\usepackage{rotating}
\usepackage{adjustbox}
\usepackage{color, colortbl}
\usepackage{amssymb}
\usepackage{subfig}
\usepackage{comment}
\usepackage{bbm}
\usepackage{amssymb}
\usepackage{graphicx}
\usepackage{titlesec}
\usepackage[most]{tcolorbox}
\usepackage{enumitem}
\usepackage{xspace}

\usepackage{makecell}

\makeatletter
\newcommand\bigDiamond{\mathop{\mathpalette\bigDi@mond\relax}}
\newcommand\bigDi@mond[2]{%
  \vcenter{\hbox{\m@th
    \scalebox{\ifx#1\displaystyle 2\else1.2\fi}{$#1\Diamond$}%
  }}%
}
\newcommand\bigLozenge{\mathop{\mathpalette\bigL@zenge\relax}}
\newcommand\bigL@zenge[2]{%
  \vcenter{\hbox{\m@th
    \scalebox{\ifx#1\displaystyle 2\else1.2\fi}{$#1\blacklozenge$}%
  }}%
}
\makeatother

\usepackage{caption}
\newcommand{\distance}{3pt}
\newcolumntype{g}{>{\columncolor{Gray}}r}
\newcolumntype{h}{>{\columncolor{Gray}}l}

\newcommand{\rltrans}{Transformer\textsubscript{Rel}\xspace}
\newcommand{\abtrans}{Transformer\textsubscript{Abs}\xspace}

\newcommand{\block}{block transformation}
\newcommand{\lineadd}{insertion / deletion transformation}
\newcommand{\linemdf}{grammatical statement transformation}
\newcommand{\tokensys}{grammatical token transformation}
\newcommand{\tokenuser}{identifier transformation}

\newcommand{\ablock}{Block transformation}
\newcommand{\alineadd}{Insertion / deletion transformation}
\newcommand{\alinemdf}{Grammatical statement transformation}
\newcommand{\atokensys}{Grammatical token transformation}
\newcommand{\atokenuser}{Identifier transformation}

\newcommand{\typeone}{$T\textsubscript{B}$}
\newcommand{\typetwo}{$T\textsubscript{ID}$}
\newcommand{\typethree}{$T\textsubscript{GS}$}
\newcommand{\typefour}{$T\textsubscript{GT}$}
\newcommand{\typefive}{$T\textsubscript{I}$}
\newcommand{\typeall}{$T\textsubscript{all}$}

\ifCLASSOPTIONcompsoc

  \usepackage[nocompress]{cite}
\else
  \usepackage{cite}
\fi

\ifCLASSINFOpdf

\else

\fi

\begin{document}

\title{A Closer Look into Transformer-Based Code Intelligence Through Code Transformation: Challenges and Opportunities}

\author{Yaoxian Li,
        Shiyi Qi,
        Cuiyun Gao,
        Yun Peng,
        David Lo,
        Zenglin Xu,
        and Michael R. Lyu

\thanks{Manuscript received April 19, 2005; revised August 26, 2015.}}

\markboth{Journal of \LaTeX\ Class Files,~Vol.~14, No.~8, August~2015}%
{Shell \MakeLowercase{\textit{et al.}}: Bare Demo of IEEEtran.cls for Computer Society Journals}

\IEEEtitleabstractindextext{%
\begin{abstract}

Transformer-based models have demonstrated state-of-the-art performance in many intelligent coding tasks such as code comment generation and code completion. Previous studies show that deep learning models are sensitive to the input variations, but few studies have systematically studied the robustness of Transformer under perturbed input code.
In this work, we empirically study the effect of semantic-preserving code transformation on the performance of Transformer. Specifically, 24 and 27 code transformation strategies are implemented for two popular programming languages, Java and Python, respectively. For facilitating analysis, the strategies are grouped into five categories: \block{}, \lineadd{}, \linemdf{}, \tokensys{}, and \tokenuser{}. Experiments on three popular code intelligence tasks, including code completion, code summarization and code search, demonstrate \lineadd{} and \tokenuser{} show the greatest impact on the performance of Transformer.
Our results also suggest that Transformer based on abstract syntax trees (ASTs) shows more robust performance than the model based on only code sequence under most code transformations. Besides,
the design of positional encoding can impact the robustness of Transformer under code transformation.
Based on our findings, we distill some
insights about the challenges and opportunities
for Transformer-based
code intelligence.

\end{abstract}

}

\maketitle

\IEEEdisplaynontitleabstractindextext

%
\IEEEpeerreviewmaketitle

\section{INTRODUCTION}
Over the past few years, deep neural networks (DNNs) have been continuously expanding their real-world applications for source code intelligence tasks
\cite{white2016deep,hu2018deep,gu2018deep,feng2020codebert}. 
Due to the format similarity between source code and text\cite{chirkova2021empirical}, Transformer\cite{vaswani2017attention}, an attention-based neural network architecture for learning textual semantics
\cite{wolf2020transformers}, is now widely used for source code representation learning
\cite{white2016deep,hu2018deep,gu2018deep,feng2020codebert}, and becomes a state-of-the-art architecture in several
code intelligence tasks, including code completion\cite{kim2021code,ciniselli2021empirical}, code summarization\cite{ahmad2020summarization,gao2021code}, and program repair\cite{berabi2021tfix}.

\begin{figure*}[htbp]
    \centering
    \includegraphics[width=0.99 \textwidth]{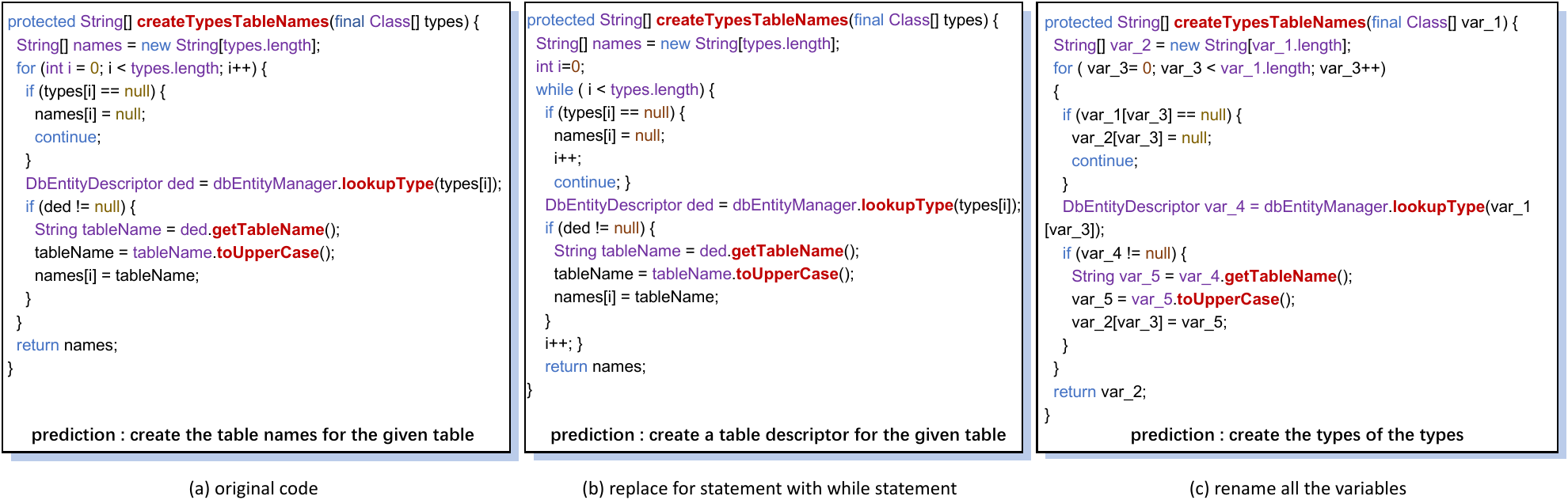}
    \setlength{\abovecaptionskip}{0.2cm} 
    \caption{Examples of semantic-preserving code transformation. The figures from left to right represent the original code, semantically equivalent programs under for-to-while transformation and variable rename transformation, respectively.}

    \label{fig:example}
\end{figure*}

Unfortunately, DNNs have demonstrated
to be quite brittle to data changes~\cite{gao2020fuzz,xie2019deephunter}. For example,
previous studies demonstrate that 
adding small perturbations to the original input
can readily trick DNNs\cite{ko2019popqorn,garg2020bae}, revealing that the DNNs are not robust to input variations\cite{carlini2017towards,jakubovitz2018improving}.
Several studies \cite{tian2018deeptest,borkar2019deepcorrect,ebrahimi2017hotflip} have been proposed to understand the robustness of DNNs under input perturbations.
Recently, Transformer\cite{vaswani2017attention}
has attracted a lot of academic attention in code intelligence tasks\cite{ahmad2020summarization,kim2021code}, but few studies have looked at its robustness when faced with perturbed code.
Given the growing number of Transformer-based code intelligence models
\cite{svyatkovskiy2020intellicode,tang2021ast,li2022setransformer,sun2020treegen}, the robustness of these models under code perturbation is of great importance.
However, developing such a robustness verification method for Transformer-based code intelligence models is challenging. Directly applying the input perturbation techniques in natural language processing (NLP) or computer vision (CV) field~\cite{zheng2020evaluating,ren2019generating} is unreasonable, since the perturbation on source code must guarantee that the changed code 
follows syntax rules. 

In this paper, we propose several semantic-preserving code transformation strategies, and analyze the impact of code transformation on the performance of Transformer.
Figure~\ref{fig:example} shows an example of semantically equivalent programs, where the code summaries are produced by the popular Transformer-based approach\cite{ahmad2020summarization}. For the code listed in Figure~\ref{fig:example}(a), we transform the \textit{if statement} to equivalent \textit{while statement}, as shown in
Figure~\ref{fig:example}(b), and conduct variable renaming, as shown in
Figure~\ref{fig:example}(c). However, the resulting summarizations of Transformer on the above three programs are radically different. Since the semantics of the original program are kept, the model should have the same prediction as the original program for the transformed programs.
This example suggests that (1) Transformer are not robust in code intelligence tasks when faced with semantic-preserving transformation, and (2) different code transformation strategies have different impacts on Transformer. Therefore, we aim at investigating
whether Transformer can maintain performance under semantic-preserving code transformation, and the impact of different transformation strategies.

In this work, we empirically study the effect of semantic-preserving code transformation on the performance of Transformer. We firstly design and implement 27 and 24 semantic-preserving transformation strategies for Java and Python languages respectively, and group them into 5 types of strategies according to the scope of influence under the transformation:
\block{}, \lineadd{}, \linemdf{}, \tokensys{}, and \tokenuser{}.
Then, we apply the transformed code on three popular code intelligence tasks: code completion (CC), code search (CS), and code summarization (CoSum). 
For studying whether involving syntax information such as Abstract Syntax Trees (ASTs) is beneficial for improving the robustness of Transformer under code transformation, we classify the Transformer-based code intelligence models into two types according to the input: Seq-based Transformer and AST-based Transformer. The seq-based Transformer only considers sequences of code tokens as input; while AST-based Transformer also involves parsed ASTs as input.

Besides, the positional encoding is an essential component in Transformer\cite{ke2020rethinking}, and has been proven effective in Transformer-based code models \cite{ahmad2020summarization, shiv2019novel, chirkova2021empirical}. Therefore, in this work, we also study the impact of different positional encoding strategies on the robustness of Transformer models under code perturbation. Specifically, two widely-used positional encoding strategies, including absolute positional encoding\cite{vaswani2017attention} and relative positional encoding\cite{shaw2018self}, are chosen for analysis.
We aim at answering the following research questions in the work.

\begin{enumerate}[label=\bfseries RQ\arabic*:,leftmargin=.5in]

     \item How do different
   code transformations impact the performance
   of Transformer? (Seq-based Transformer)
   
   \item Is AST helpful for reducing
   the impact of code transformations on the performance of Transformer? (AST-based Transformer)

\end{enumerate}\textbf{}

During answering each research question, we also consider different positional encoding strategies. We achieve some findings and summarize the key findings as below.

\begin{itemize}

    \item Code transformations such as \lineadd{} and \tokenuser{} present greatest impact on the performance of both seq-based and AST-based Transformers.

    \item Transformer based on ASTs shows more robust performance than the model based on only code sequence under most code transformations.

    \item The relative position encoding can improve the robustness of seq-based Transformer under most code transformations, but has no much benefit for the robustness of AST-based Transformer.

\end{itemize}

Based on the findings, we derive some insights about the challenges and opportunities that would benefit future research. For example, future work is expected to better exploit ASTs to
boost the robustness of Transformer models under code transformation.
Besides, we also encourage future work to explore more effective attention approach or engage additional external knowledge to eliminate the distraction of \lineadd{}.
Furthermore, future work should eliminate the impact of identifiers during code representation learning, instead of relying on the semantics of identifiers.

The main contributions of this paper are summarized as follows:

\begin{itemize}

    \item We empirically study the effect of semantic-preserving code transformation on the performance of Transformer for three popular code intelligence tasks.
    
    \item We design and implement
    27 and 24 code transformation strategies for Java and Python languages, respectively.

    \item We study how different aspects can impact the performance of Transformer, including the input and positional encoding strategy.
    
    \item We achieve some findings and implications that would benefit future research in the robustness of Transformer-based code intelligence tasks.
    
\end{itemize}

The rest of this paper is organized as follows. We present the background of Transformer and code intelligence tasks in Section~\ref{sec:backgroud}. The technical details of our code transformation strategies are presented in Section~\ref{sec:approach}. The evaluation and study design are shown in Section~\ref{sec:experiment}.Then we present the experimental results and potential findings in Section~\ref{sec:result}. Based on the findings, we conclude some implications and future directions in Section~\ref{sec:discuss}.
We discuss threats to validity in Section~\ref{sec:therats}. Finally, we give the review of the literature related to our research in Section~\ref{sec:related} and conclude the work in Section~\ref{sec:conclusion}, respectively.

\section{BACKGROUND}
\label{sec:backgroud}

\subsection{Transformer and positional encoding}

Transformer employs the typical encoder-decoder structure\cite{vaswani2017attention}, and is composed of stacked Transformer blocks. Each block contains a multi-head self-attention sub-layer followed by a fully connected positional-wise feed-forward network sub-layer. The sub-layers are connected by residual connections\cite{he2016deep} and layer normalization\cite{ba2016layer}. Different from the encoder,
the decoder
has
attention sub-layers that use the key and value matrices from the encoder. Positional encoding is an essential component in Transformer\cite{ke2020rethinking}, and has proven effective in code intelligence tasks \cite{chirkova2021empirical}. We then introduce the two popular positional encoding strategies, including absolute positional encoding \cite{vaswani2017attention} and relative positional encoding \cite{shaw2018self}, as below.

\textbf{Absolute positional encoding.}
The original Transformer is supplemented by positional encoding to accommodate for the input's sequential nature. It transposes the sequence of input vectors $\mathbf{X}=(x_1,x_2,...,x_n)$ into the sequence of output vectors $\mathbf{Z}=(z_1,z_2,...,z_n)$, where $x_i$, $z_i\in {R^{d_{model}}}$. When doing self attention, Transformer first projects the input vector $\mathbf{X}$ into three vectors: the query $Q$, key $K$ and value $V$ by trainable parameters $W^{Q}$, $W^{K}$, $W^{V}$. The attention weight is calculated using dot product and softmax function. The output vector is the weighted sum of the value vector:\par
\begin{equation}
    e_{ij}=\frac{(x_iW^Q)(x_jW^K)^T}{\sqrt{d}}, \label{eq:1}
\end{equation}
\begin{equation}
    \alpha_{ij}= \frac{exp\;e_{ij}}{\sum_{k=1}^{n} exp\;e_{ij}}, \label{eq:2}
\end{equation}
\begin{equation}
    z_i=\sum\limits_{j=1}^{n}\alpha_{ij}(x_{j}W^{V}), \label{eq:3}
\end{equation}
where $d$ is the dimension of each vector, and is used to scale the dot product.

\textbf{Relative positional encoding.}
To encode the pairwise positional relationships between input elements, Shaw et al. \cite{shaw2018self} propose the relative position encoding which models the relation of two elements through their distance in the input sequence.
Formally, the relative position embedding between input element $x_i$ and $x_j$ is represented as $a_{ij}^V$,$a_{ij}^K$ $\in$ $R^{d}$. In this way, the self attention calculated in Equ. (~\ref{eq:1}) and Equ. (~\ref{eq:3}) can be rewritten as:
\begin{equation}
    e_{ij}=\frac{(x_iW^Q)(x_jW^K+a_{ij}^K)^T}{\sqrt{d_z}}, \label{eq:4}
\end{equation}
\begin{equation}
   z_i=\sum\limits_{j=1}^{n}\alpha_{ij}(x_{j}W^{V}+a_{ij}^V). \label{eq:5}
\end{equation}
Relative position representations take the relative distance into calculating attention rather than absolute position, which perform more effectively and flexibly.

\subsection{Code intelligence task}

\textbf{Code completion task.} Code completion is commonly used in modern integrated development environments (IDEs) for facilitating programming~\cite{svyatkovskiy2021fast}. 
Developers use the code completion technique to predict expected code elements, such as class names and methods, based on given code surrounding the point of prediction\cite{ciniselli2021empirical}. 
Common code completion techniques include token-level completion and statement-level completion\cite{izadi2022codefill}. In our experiment, we focus on the token-level completion, and the task is to predict the next code token ($n_i$) based on the previous code tokens [$n_0$,$n_1$,...,$n_{i-1}$].

\textbf{Code summarization task.} The task of code summarization is to generate a natural language summary (e.g., a doc-string) for given source code\cite{feng2020codebert,zhang2020retrieval}. Code summary can help developers to understand the function and purpose of code without requiring the developers to read the code itself, which can save them time from comprehending the details of that code\cite{leclair2020improved}. 
For a dataset containing a set of programs $C$ and targeted summaries $S$, the task of code summarization is to generate the summary consisting of a sequence of token $\tilde{s} = (s_0,s_1,...,s_{m})$ by maximizing the conditional likelihood
$\tilde{s} = \mathop{\arg\max}_s P(s|c)$ for the given code $c = (c_0,c_1,...,c_{n})$ from $C$, where $s$ is the corresponding summary in $S$.

\textbf{Code search task.} The goal of code search is to find the most semantically-related code from a collection of code based on a given natural language query \cite{husain2019codesearchnet}. 
In our experiment, we focus on neural code search\cite{cambronero2019deep,gu2021multimodal}, which learns the joint embeddings of natural-language query and code snippet
\cite{fang2021self}. 
The task of neural code search is to return the expected ranking order of code snippets for the given natural language.

\section{CODE TRANSFORMATION}
\label{sec:approach}

\begin{figure*}
    \centering
    \includegraphics[width=0.95 \textwidth]{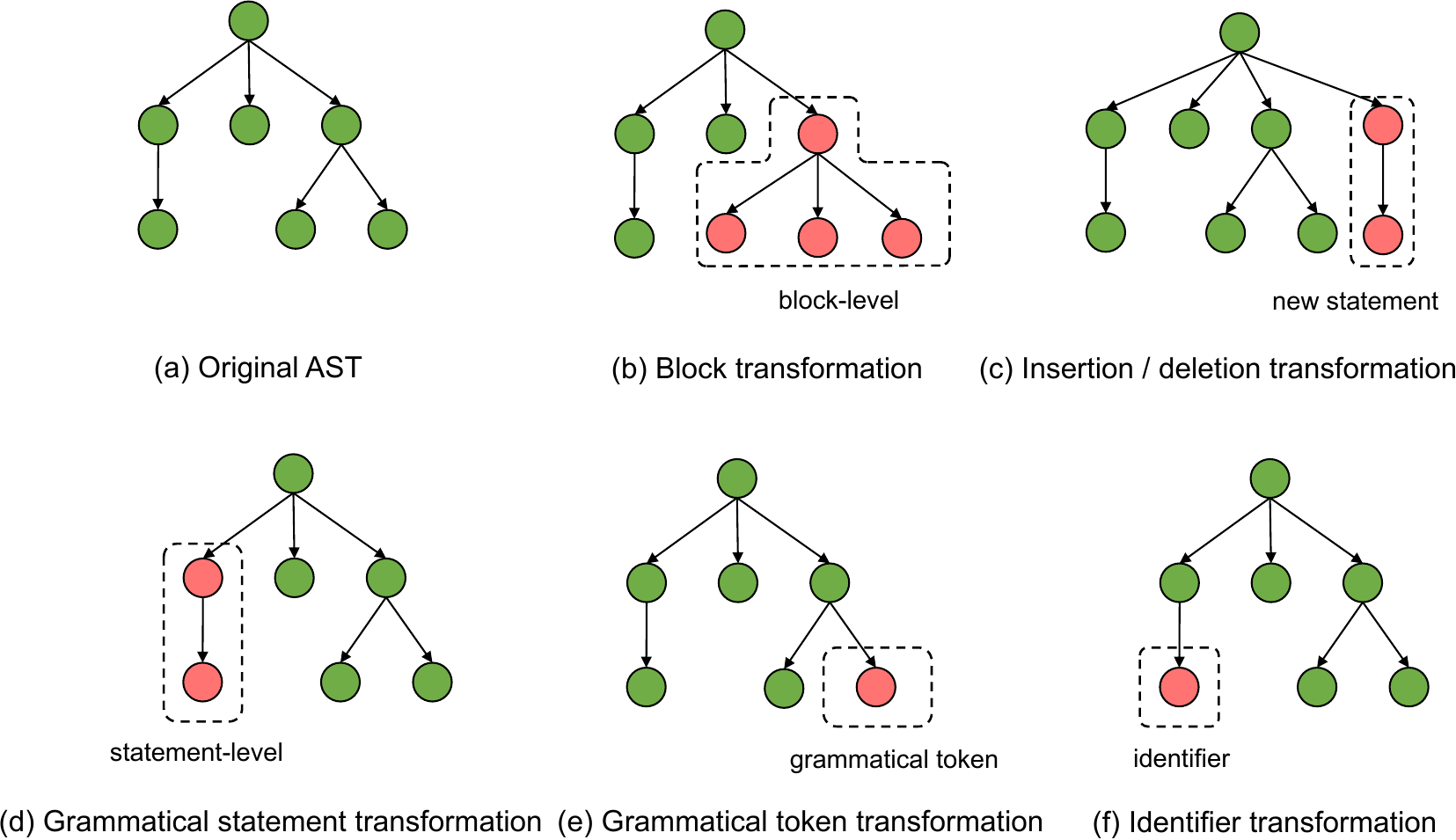}
    \setlength{\abovecaptionskip}{0.2cm} 
    \caption{ Example of the code transformation. Figure (a) is an AST schematic, and figures (b)-(f) illustrate the different structure changes at the AST level with different code transformations.} 
    
    \label{fig:code_trans_example}
\end{figure*}

\begin{table*}[htbp]
\caption{Semantic-preserving code transformation in our experiment.}
\begin{adjustbox}{width=1\textwidth}
\begin{tabular}{lllll}
\toprule
No. & Transformation strategy & Description of transformation & Java & Python\\
\midrule

\multicolumn{4}{c}{\centerline{\textbf{Block transformation}}} \\
\midrule
B-1 & For statement  & replace the for statement by equivalent while statement & \Checkmark & \Checkmark
\\
B-2 & While statement  & replace the while statement by equivalent for statement &\Checkmark & \XSolidBrush \\
B-3 & Elseif to if else & \begin{tabular}[c]{@{}l@{}}  convert enseif to else if, for example, if(x==1)\{ \}else\{if(x==2)\{ \} \} becomes if (x==1)\{ \}\\else if (x==2)\{ \} \end{tabular} & \Checkmark & \Checkmark \\
B-4 & Else if to elseif & \begin{tabular}[c]{@{}l@{}} convert else if to elseif, for example, if(x==1)\{ \}else if (x==2) \{ \} becomes if (x==1) \{ \} \\ else \{ if(x==2)\{ \} \} \end{tabular}    & \Checkmark & \Checkmark \\
B-5 & If to else & \begin{tabular}[c]{@{}l@{}} use logical not operator to change the condition of if statement and exchange the \\ block of if and else, for example, if(x==0)\{ block1 \}else\{ block2 \} becomes if !(x==0)\\ \{ block2 \} else \{ block1 \}\end{tabular} & \Checkmark & \Checkmark\\
B-6 & Change if statement & \begin{tabular}[c]{@{}l@{}} 
if the condition of if statement has logical operator(e.g, \&\& ), it will split the if statement \end{tabular} & \Checkmark & \Checkmark \\
B-7 & Create new function & \begin{tabular}[c]{@{}l@{}} move the variable initialization statement to generate a new function, and then call \\ the function, for example, z = x+y  becomes def func(x,y): return x+y  z=func(x,y) \end{tabular} & \XSolidBrush & \Checkmark \\
\midrule

\multicolumn{4}{c}{\centerline{\textbf{Insertion / deletion transformation. }
}}.  \\
\midrule
ID-1 & Add comments  & insert comments not related to the source code  & \Checkmark & \Checkmark \\
ID-2 & Add junk code  & add code that not related to the source code  & \Checkmark & \Checkmark \\
ID-3 & Add return statement  & add a return statement at the end of the source code that returns the default value  & \Checkmark & \Checkmark \\
ID-4 & Import library & import libraries unrelated to source code & \Checkmark & \Checkmark \\
ID-5 & Delete comment & remove all the comments from source code  & \Checkmark & \Checkmark \\
ID-6 & Delete print & replace the print statement by empty statement, for example, replace print by pass  & \Checkmark  & \Checkmark \\
ID-7 & Remove unused variable & remove the variable declaration statement if the variable is never used  & \Checkmark & \Checkmark \\
\midrule

\multicolumn{4}{c}{\centerline{\textbf{Grammatical statement transformation}}} \\
\midrule

GS-1 & Change return statement & \begin{tabular}[c]{@{}l@{}}
if the return statement returns an integer literal, we will declare a variable, and \\return the variable\end{tabular}  & \Checkmark & \Checkmark \\
GS-2 & For move in variable declaration & \begin{tabular}[c]{@{}l@{}} move the variable declaration into for statement. For instance, int i; for(i=0;i<10;i++)\\ becomes for(int i=0;i<10;i++) \end{tabular} & \Checkmark & \XSolidBrush \\ 
GS-3 & For move out variable declaration & move the variable declaration out of for statement & \Checkmark & \XSolidBrush \\
GS-4 & Change variable declaration & \begin{tabular}[c]{@{}l@{}} split the variable declaration and initialization, for example, int i=0 becomes int i i=0 \end{tabular} & \Checkmark & \XSolidBrush \\
GS-5 & Add logical operator & \begin{tabular}[c]{@{}l@{}} add the logical not operator and use opposite comparison operator, for example, x$<$y \\ becomes !(y$>=$x) \end{tabular} & \Checkmark & \Checkmark \\
GS-6 & Change comparison operator & use the opposite comparison operator, for example, x$<$y becomes y$>$x & \Checkmark & \Checkmark \\
GS-7 & Change argument assignment operator & \begin{tabular}[c]{@{}l@{}}  change the argument assignment operator to assignment operator, for example, x+=1 \\ becomes x=x+1\end{tabular} & \Checkmark & \Checkmark \\
GS-8 & Change the unary operator & change the increment or decrease operator. For instance, x++ becomes x=x+1 & \Checkmark & \XSolidBrush \\
GS-9 & Add curly bracket & add a curly brace to a single statement and then generate a new compound statement  & \Checkmark & \XSolidBrush \\
GS-10 & Delete curly bracket & \begin{tabular}[c]{@{}l@{}}if the compound statement has only a single statement, delete curly of the compound \\statement \end{tabular}  & \Checkmark & \XSolidBrush \\
\midrule

\multicolumn{4}{c}{\centerline{\textbf{Grammatical token transformation}.}} \\
\hline

GT-1 & Bool to int  &  replace True or False by 1 or 0 & \XSolidBrush & \Checkmark \\
GT-2 & Int to bool  & replace 1 or 0 by True or False  & \XSolidBrush & \Checkmark \\
GT-3 & Upper integral type  & replace the integral type by a higher type, for example, replace int by long & \Checkmark & \Checkmark \\
GT-4 & Upper floating type  & replace integral type by floating type or replace float by double & \Checkmark & \Checkmark \\

GT-5 & Change input API & change the API for reading input & \XSolidBrush & \Checkmark \\
GT-6 & Change output API & change the API for writing output & \Checkmark & \Checkmark \\
\midrule

\multicolumn{4}{c}{\centerline{\textbf{Identifier transformation}}} \\
\midrule
I-1 & Function rename & rename the function name and class name & \Checkmark & \Checkmark \\
I-2 & Variable rename & rename the variable name & \Checkmark & \Checkmark \\

\bottomrule
\end{tabular}
\end{adjustbox}
\label{tab:codetrans}%
\end{table*}

To verify the robustness of Transformer under input perturbations, we have implemented 24 and 27 semantic-preserving code transformation strategies for Python and Java languages, respectively.
The
semantic-preserving code transformation is implemented on AST, and consists of three phases: (1) we parse the source code into its AST
using the standard compiler tool (e.g., tree-sitter\footnote{https://tree-sitter.github.io/tree-sitter/} in our experiments); 
(2) we directly modify the structure of AST to the target code formation; (3) we convert the modified AST to a transformed source code. This process also needs to
make sure the transformed code can be compilable and executable.

Table~\ref{tab:codetrans} presents all the transformation strategies and corresponding descriptions. To conduct a thorough investigation of the impact of transformed code on Transformer, we classify the code transformation strategies into five types according to the scope of influence under the transformation:
\begin{itemize}
    \item \ablock;
    \item \alineadd;
    \item \alinemdf;
    \item \atokensys;
    \item \atokenuser.
\end{itemize}

\noindent
For instance, the \textit{bool to int transformation} converts the Boolean value from True/False to 1/0 and only affects the changed Boolean token, hence it belongs to \textit{\tokensys}.

\textbf{\ablock{} (denoted as $T\textsubscript{B}$).} This type refers to the code transformation that impacts the code at block level, as shown in Figure~\ref{fig:code_trans_example} (b). The example illustrated in Figure~\ref{fig:example} (b) is a block transformation, where the \textit{if-statement} is transformed to the equivalent \textit{while-statement}. The type contains seven code transformation strategies in total.

\textbf{\alineadd{} (denoted as $T\textsubscript{ID}$).} This type of transformation is implemented
at the statement level. The transformation generally adds
new statements or deleting existing ones without impacting
other statements in the program. 
During implementation, only sub-trees are added or deleted at the AST level, as depicted in Figure~\ref{fig:code_trans_example} (c).
For example, the \textit{remove unused variable transformation} will remove
the variable declaration statement if the variable is never used again, and such change will not affect other statements in the program.
This type consists of 7 code transformation strategies.

\textbf{\alinemdf{} (denoted as $T\textsubscript{GS}$).} This type of transformation is also operated at the statement level. Different from $T\textsubscript{ID}$, $T\textsubscript{GS}$ changes the statements in the original code, as depicted in Figure~\ref{fig:code_trans_example} (d). For example, the \textit{change the unary operator} replaces the statement \textit{i++} by \textit{i=i+1}.
This type has 10 code transformation strategies.

\textbf{\atokensys{} (denoted as $T\textsubscript{GT}$).} This type of transformation changes the original code at the token level, and includes six code transformation strategies. As illustrated in Figure~\ref{fig:code_trans_example} (e), the transformation only affects the type and value of associated AST nodes, with the structure unchanged. Note that the type of transformation does not involve identifiers.
For instance, the \textit{bool to int transformation} converts the Boolean operator from True/False to 1/0.

\textbf{\atokenuser{} (denoted as $T\textsubscript{I}$).} This type of transformation is also implemented at token level but manly operates on identifiers, as illustrated in Figure~\ref{fig:code_trans_example} (f). The type includes two transformation strategies, including \textit{function rename transformation} and \textit{variable rename transformation}. For example, the the \textit{variable rename transformation} renames the identifiers (variable name) with placeholders such as var1 and var2.
Additionally, some strategies cannot be implemented for both languages considering the language-specific characteristics.
For example, the Python language does not support the increment and decrement operators, so the \textit{change the unary operator transformation} strategy is only allowed
in Java. 
The Java
language treats Boolean as a unique data type with two distinct values: True and False, so the \textit{bool to int transformation} strategy is only applicable for Python.

\section{EXPERIMENTAL SETUP}
\label{sec:experiment}

\subsection{Datasets and pre-processing}

Following the prior work\cite{parvez2021retrieval}, we choose the Java and Python datasets from CodeSearchNet~\cite{husain2019codesearchnet} for evaluation.
CodeSearchNet
is a collection of large datasets with code-document pairs
from open source projects on GitHub, which is widely-used in code intelligence tasks\cite{mastropaolo2021studying,lin2021traceability}. Detailed data statistics are illustrated in Table~\ref{tab:stats}. The subject data consist of 165K / 5K / 11K training / validation / test code snippets for Java and 252K / 14K / 15K for Python, respectively. For facilitating analysis, we parse the code snippets into ASTs, and traverse the ASTs into sequences of tokens as input in depth-first-search order following~\cite{chirkova2021empirical}.

\begin{table}[ht!]
  \centering
  \caption{Statistics of experimental data.
  }
    \begin{tabular}{lrrr}
    \toprule
    \textcolor[rgb]{ .141,  .161,  .184}{Language} & \multicolumn{1}{l}{\textcolor[rgb]{ .141,  .161,  .184}{Train}} & \multicolumn{1}{l}{\textcolor[rgb]{ .141,  .161,  .184}{Valid}} & \multicolumn{1}{l}{\textcolor[rgb]{ .141,  .161,  .184}{Test}} \\
    \midrule
    \textcolor[rgb]{ .141,  .161,  .184}{Python} & \textcolor[rgb]{ .141,  .161,  .184}{251,820} & \textcolor[rgb]{ .141,  .161,  .184}{13,914} & \textcolor[rgb]{ .141,  .161,  .184}{14,918} \\
    \textcolor[rgb]{ .141,  .161,  .184}{Java} & \textcolor[rgb]{ .141,  .161,  .184}{164,923} & \textcolor[rgb]{ .141,  .161,  .184}{5,183} & \textcolor[rgb]{ .141,  .161,  .184}{10,955} \\
    \bottomrule
    \end{tabular}%
  \label{tab:stats}%
\end{table}%

\subsection{Implementation}
In this work, we consider three 
code intelligence tasks: code completion, code summarization, and code search. We elaborate on the detailed implementation of the three tasks in the following.

\textbf{Code completion.}
We use the setup, metrics and implementation of Transformer according to\cite{kim2021code}. Besides, we split code sequence whose length is over 500 following\cite{chirkova2021empirical}.

\textbf{Code summarization.} 
In our experiments, we select \cite{ahmad2020summarization} as Transformer implementation except that we do not split sub-token following Chirkova et al.~\cite{chirkova2021empirical}.

\textbf{Code search.} 
In our experiments, we implement a Transformer architecture for code search task based on \cite{papathomas2022semantic}.
We process the dataset following the strategy of Evangelos et. al \cite{papathomas2022semantic}. For example, we filter the non-ASCII tokens and replace symbols by their English names (e.g., the symbol $+$ in code token is replace by \texttt{addoperator}).

\textbf{Hyper-parameters.} The hyper-parameters setting in our experiments follows Transformer implementations\cite{kim2021code,ahmad2020summarization,papathomas2022semantic} and we list the major hyper-parameters for code completion, code summarization, and code search tasks. Our Transformer models include 6 layers, 6 heads with the layers of our models to be 512. 
We set the maximum distance of relative attention to 32 for all tasks. We train Transformers using Adam with a starting learning rate of 0.0001 and a batch size of 32 / 32 / 128 with the epoch number being 15 / 20 / 100 for code completion, summarization, and search respectively.
We train all models on 4 GPUs of Nvidia Tesla V100 with 32G of memory.

\textbf{Evaluation set.} As illustrated in Table~\ref{tab:codetrans}, not all the transformation strategies are applicable for both programming languages. For example, the \textit{bool to int} strategy is only allowed in Python. During analyzing the impact of each code transformation strategy on models' performance, we only evaluate on the transformable code instead of all the code in the test set. Besides, to minimize the performance bias, we run each experiment for three times and report the average results.

\subsection{Evaluation metrics}

\subsubsection{Code summarization}
We evaluate the source code summarization performance using three metrics: BLEU, METEOR and ROUGE-L.

\textbf{BLEU} is a widely-used metric in natural language processing and software engineering fields to evaluate the quality of generated texts, e.g., machine translation, code comment generation, and code commit message generation~\cite{papineni2002bleu}. It computes the frequencies of the co-occurrence of n-grams between the ground truth $\hat{y}$ and the generated sequence $y$ to judge the similarity:
$$
\mathrm{\text{BLEU-N}}=\mathrm{b(y,\hat{y})} \cdot \exp \left(\sum_{n=1}^{N} \beta_{n} \log p_{n}(y,\hat{y})\right),
$$
where $\mathrm{b(y,\hat{y})}$ indicates the brevity penalty, and $p_{n}(y,\hat{y})$ and $\beta_{n}$ represent the geometric average of the modified n-gram precision and the weighting parameter, respectively.

\textbf{ROUGE-L} is commonly used in natural language translation~\cite{lin2004rouge}, and is a F-measure based on the Longest Common Subsequence (LCS) between candidate and target sequences, where the LCS is a set of words appearing in the two sequences in same order. 
$$
ROUGE\text{-}L=\frac{\left(1+\beta^{2}\right) R_{l c s} P_{l c s}}{R_{l c s}+\beta^{2} P_{l c s}},
$$
where $R_{l c s}=\frac{L C S(X, Y)}{len(Y)}$ and $P_{l c s}=\frac{L C S(X, Y)}{len(X)}$. $X$ and $Y$ denote candidate sequence and reference sequence, respectively. $L C S(X, Y)$ represents the length of the longest common sub-sequence between $X$ and $Y$.

\textbf{Meteor} is an evaluation metric proposed based on BLEU\cite{banerjee2005meteor}. It introduces synonym, stem, and other information to replace the precise matching in BLEU, and strengthens the role of recall in automatic evaluation.

\subsubsection{Code search and code completion}
For code search and code completion tasks, we use MRR\cite{radev2002evaluating} as the evaluation metric. MRR is the average of the reciprocal rank of results of a set of queries. The reciprocal rank of a query is the inverse of the rank of the first hit result. 
$$
MRR = \frac{1}{N} \sum_{n=1}^{N} (\frac{1}{rank_{i}}),
$$
where $N$ is the total number of targets (tokens for code completion and code snippet for code search) in data pool and $rank_{i}$ represents the position of the $i$-th true target in the ranking results.

\section{RESEARCH QUESTIONS AND RESULT ANALYSIS}
\label{sec:result}

Our experimental study aims to answer the following research questions:

\begin{enumerate}[label=\bfseries RQ\arabic*:,leftmargin=.5in]
   \item How do different
   code transformations impact the performance
   of Transformer? (Seq-based Transformer)
   
   \item Is AST helpful for reducing
   the impact of code transformations on the performance of Transformer? (AST-based Transformer)


\end{enumerate}\textbf{}

RQ1 aims at discovering which types of code transformation show greatest impact on the robustness of Transformer. RQ2 aims at analyzing whether AST is beneficial for improving the robustness of Transformer under different code transformations. During answering both RQs, we also consider the impact of different positional encoding strategies.

\subsection{Answer to RQ1: Impact on seq-based Transformer}
\label{sec:q1}
In this section, we compare the performance of Transformer before and after different code transformations for three different tasks.
Table~\ref{tab:cc_text}, Table~\ref{tab:cs_text} and Table~\ref{tab:csm_text} present the overall results of code completion, code search and code summarization, respectively. 

]

\subsubsection{Different types of code transformation on code sequence} 
\label{sec:q1_type}
In the section, we analyze the effects of different types of code transformation on the performance of seq-based Transformer.
We observe that seq-based Transformer's performance is affected to varying degrees by different types of code transformations. 
We elaborate on the detailed impact of different types of code transformation in the following.

\textbf{\ablock{} (\typeone).} 
As shown in Table~\ref{tab:cc_text}-\ref{tab:csm_text}, we observe that Transformer demonstrates robust performance under \textit{\block{}} on all code intelligence tasks.
For example, the MRR values for the code completion task just decrease
by 0.81\% and 0.65\% for Java and Python, respectively (seen in Table~\ref{tab:cc_text}).

\textbf{\alineadd{} (\typetwo).} 
From Table~\ref{tab:cc_text}-\ref{tab:csm_text}, we observe that
\textit{\lineadd} 
has a substantial impact on Transformer on all code intelligence tasks. For example, the decrease of Transformer on the code search task is from 23.34\% to 26.77\% (seen in Table~\ref{tab:cs_text}). When generating Java's code summary, the BLEU, ROUGE-L, and METEOR values decrease by 5.31\%, 7.85\%, and 10.84\%, respectively (seen in Table~\ref{tab:csm_text}).

\begin{table}[htbp]
  \centering
  \caption{Results of code transformation on the performance of seq-based Transformer for the code completion task.
  Column ``Pos.'' represents the position encoding strategy,
  while ``abs.'' and ``rel.'' represent the absolute/relative position encoding, respectively. Column ``w. t.'' and ``w/o t.'' represent the results 
  with and without code transformation. Different rows represent the results of different
  types of code transformation, while the bottom portion \typeall{} presents the average results. The \textcolor{red}{red} color indicates the degree of decrease.
  }
  \begin{adjustbox}{width=.49\textwidth}
    \begin{tabular}{cc|ccc|ccc}
    \toprule
    \multicolumn{2}{c|}{} & \multicolumn{3}{c|}{MRR (Java)} & \multicolumn{3}{c}{MRR (Python)} \\
    Type  & Pos.  & w/o t. & w. t. & imp. (\%) & w/o t. & w. t. & imp. (\%) \\
    \midrule
    \multirow{2}[2]{*}{\typeone} & abs.  & 0.7243 & 0.7185 & \color{red}$\downarrow$ \textcolor[rgb]{ 1,  0,  0}{0.81} & 0.6581 & 0.6539 & \color{red}$\downarrow$ \textcolor[rgb]{ 1,  0,  0}{0.65} \\
          & rel.  & 0.7343 & 0.7290 & \color{red}$\downarrow$ \textcolor[rgb]{ 1,  0,  0}{0.72} & 0.6708 & 0.6672 & \color{red}$\downarrow$ \textcolor[rgb]{ 1,  0,  0}{0.54} \\
    \midrule
    \multirow{2}[1]{*}{\typetwo} & abs.  & 0.5925 & 0.5295 & \color{red}$\downarrow$ \textcolor[rgb]{ 1,  0,  0}{10.62} & 0.5483 & 0.4801 & \color{red}$\downarrow$ \textcolor[rgb]{ 1,  0,  0}{12.45} \\
          & rel.  & 0.5999 & 0.5677 & \color{red}$\downarrow$ \textcolor[rgb]{ 1,  0,  0}{5.37} & 0.5580 & 0.5376 & \color{red}$\downarrow$ \textcolor[rgb]{ 1,  0,  0}{3.65} \\
    \midrule
    \multirow{2}[1]{*}{\typethree} & abs.  & 0.7201 & 0.7082 & \color{red}$\downarrow$ \textcolor[rgb]{ 1,  0,  0}{1.66} & 0.6488 & 0.6377 & \color{red}$\downarrow$ \textcolor[rgb]{ 1,  0,  0}{1.72} \\
          & rel.  & 0.7304 & 0.7186 & \color{red}$\downarrow$ \textcolor[rgb]{ 1,  0,  0}{1.61} & 0.6614 & 0.6502 & \color{red}$\downarrow$ \textcolor[rgb]{ 1,  0,  0}{1.69} \\
    \midrule
    \multirow{2}[1]{*}{\typefour} & abs.  & 0.7180 & 0.7107 & \color{red}$\downarrow$ \textcolor[rgb]{ 1,  0,  0}{1.02} & 0.6509 & 0.6301 & \color{red}$\downarrow$ \textcolor[rgb]{ 1,  0,  0}{3.19} \\
          & rel.  & 0.7277 & 0.7191 & \color{red}$\downarrow$ \textcolor[rgb]{ 1,  0,  0}{1.18} & 0.6637 & 0.6565 & \color{red}$\downarrow$ \textcolor[rgb]{ 1,  0,  0}{1.08} \\
    \midrule
    \multirow{2}[1]{*}{\typefive} & abs.  & 0.7169 & 0.6485 & \color{red}$\downarrow$ \textcolor[rgb]{ 1,  0,  0}{9.54} & 0.6594 & 0.6019 & \color{red}$\downarrow$ \textcolor[rgb]{ 1,  0,  0}{8.73} \\
          & rel.  & 0.7255 & 0.6544 & \color{red}$\downarrow$ \textcolor[rgb]{ 1,  0,  0}{9.81} & 0.6703 & 0.6136 & \color{red}$\downarrow$ \textcolor[rgb]{ 1,  0,  0}{8.46} \\
    \midrule
    \multirow{2}[2]{*}{\typeall} & abs.  & 0.6944 & 0.6631 & \color{red}$\downarrow$ \textcolor[rgb]{ 1,  0,  0}{4.50} & 0.6331 & 0.6007 & \color{red}$\downarrow$ \textcolor[rgb]{ 1,  0,  0}{5.12} \\
          & rel.  & 0.7036 & 0.6777 & \color{red}$\downarrow$ \textcolor[rgb]{ 1,  0,  0}{3.67} & 0.6448 & 0.6250 & \color{red}$\downarrow$ \textcolor[rgb]{ 1,  0,  0}{3.07} \\
    \bottomrule
    \end{tabular}%
    \end{adjustbox}
  \label{tab:cc_text}%
\end{table}%

\begin{table}[htbp]
  \centering
  \caption{Results of code transformation on the performance of seq-based Transformer for the code search task. The \textcolor{red}{red} and \textcolor[rgb]{ 0,  .69,  .314}{green} colors indicate the degree of decrease and increase, respectively.}
  \begin{adjustbox}{width=.49\textwidth}
    \begin{tabular}{cc|ccc|ccc}
    \toprule
    \multicolumn{2}{c|}{} & \multicolumn{3}{c|}{MRR (Java)} & \multicolumn{3}{c}{MRR (Python)} \\
    Type  & Pos.  & w/o. t. & w. t. & imp. (\%) & w/o. t. & w. t. & imp. (\%) \\
    \midrule
    \multirow{2}[2]{*}{\typeone} & abs.  & 0.3630 & 0.3611 & \color{red}$\downarrow$ \textcolor[rgb]{ 1,  0,  0}{0.52} & 0.3853 & 0.3834 & \color{red}$\downarrow$ \textcolor[rgb]{ 1,  0,  0}{0.49} \\
          & rel.  & 0.3601 & 0.3614 & \color[rgb]{ 0,  .69,  .314}$\uparrow$ \textcolor[rgb]{ 0,  .69,  .314}{0.35} & 0.3808 & 0.3779 & \color{red}$\downarrow$ \textcolor[rgb]{ 1,  0,  0}{0.76} \\
    \midrule
    \multirow{2}[1]{*}{\typetwo} & abs.  & 0.3249 & 0.2379 & \color{red}$\downarrow$ \textcolor[rgb]{ 1,  0,  0}{26.77} & 0.2802 & 0.2148 & \color{red}$\downarrow$ \textcolor[rgb]{ 1,  0,  0}{23.34} \\
          & rel.  & 0.3452 & 0.2599 & \color{red}$\downarrow$ \textcolor[rgb]{ 1,  0,  0}{24.71} & 0.2730 & 0.2080 & \color{red}$\downarrow$ \textcolor[rgb]{ 1,  0,  0}{23.82} \\
    \midrule
    \multirow{2}[1]{*}{\typethree} & abs.  & 0.4208 & 0.4221 & \color[rgb]{ 0,  .69,  .314}$\uparrow$ \textcolor[rgb]{ 0,  .69,  .314}{0.30} & 0.3891 & 0.3857 & \color{red}$\downarrow$ \textcolor[rgb]{ 1,  0,  0}{0.87} \\
          & rel.  & 0.4350 & 0.4367 & \color[rgb]{ 0,  .69,  .314}$\uparrow$ \textcolor[rgb]{ 0,  .69,  .314}{0.39} & 0.3861 & 0.3838 & \color{red}$\downarrow$ \textcolor[rgb]{ 1,  0,  0}{0.60} \\
    \midrule
    \multirow{2}[1]{*}{\typefour} & abs.  & 0.4403 & 0.4397 & \color{red}$\downarrow$ \textcolor[rgb]{ 1,  0,  0}{0.14} & 0.3761 & 0.3368 & \color{red}$\downarrow$ \textcolor[rgb]{ 1,  0,  0}{10.46} \\
          & rel.  & 0.4496 & 0.4485 & \color{red}$\downarrow$ \textcolor[rgb]{ 1,  0,  0}{0.26} & 0.3629 & 0.3248 & \color{red}$\downarrow$ \textcolor[rgb]{ 1,  0,  0}{10.52} \\
    \midrule  
    \multirow{2}[1]{*}{\typefive} & abs.  & 0.4363 & 0.2521 & \color{red}$\downarrow$ \textcolor[rgb]{ 1,  0,  0}{42.21} & 0.4093 & 0.2529 & \color{red}$\downarrow$ \textcolor[rgb]{ 1,  0,  0}{38.22} \\
          & rel.  & 0.4428 & 0.2586 & \color{red}$\downarrow$ \textcolor[rgb]{ 1,  0,  0}{41.59} & 0.4050 & 0.2516 & \color{red}$\downarrow$ \textcolor[rgb]{ 1,  0,  0}{37.88} \\
    \midrule
    \multirow{2}[2]{*}{\typeall} & abs.  & 0.3971 & 0.3426 & \color{red}$\downarrow$ \textcolor[rgb]{ 1,  0,  0}{13.73} & 0.3680 & 0.3147 & \color{red}$\downarrow$ \textcolor[rgb]{ 1,  0,  0}{14.47} \\
          & rel.  & 0.4066 & 0.3530 & \color{red}$\downarrow$ \textcolor[rgb]{ 1,  0,  0}{13.17} & 0.3616 & 0.3092 & \color{red}$\downarrow$ \textcolor[rgb]{ 1,  0,  0}{14.47} \\
    \bottomrule
    \end{tabular}%
    \end{adjustbox}
  \label{tab:cs_text}%
\end{table}%

\begin{table*}[htbp]
  \centering
  \caption{Results of code transformation on the performance of seq-based Transformer for the code summarization task. The above part presents the results of Transformer with absolute/relative position encoding strategies (``abs.'' and ``rel.'') for Java, and the below part presents results for Python. The \textcolor{red}{red} and \textcolor[rgb]{ 0,  .69,  .314}{green} color indicate the degree of decrease and increase, respectively.}
\scalebox{0.95}{
    \begin{tabular}{cc|ccc|ccc|ccc}
    \toprule
    \multicolumn{2}{c|}{Java} & \multicolumn{3}{c|}{BLEU} & \multicolumn{3}{c|}{ROUGE-L} & \multicolumn{3}{c}{METEOR} \\
    Type  & Pos.  & w/o. t. & w. t  & Imp. (\%) & w/o. t. & w. t  & Imp. (\%) & w/o. t. & w. t  & Imp. (\%) \\
    \midrule
    \multirow{2}[2]{*}{\typeone} & abs.  & 11.73 & 11.83 & \color[rgb]{ 0,  .69,  .314}$\uparrow$ \textcolor[rgb]{ 0,  .69,  .314}{0.81} & 21.82 & 21.93 & \color[rgb]{ 0,  .69,  .314}$\uparrow$ \textcolor[rgb]{ 0,  .69,  .314}{0.50} & 6.08  & 6.30  & \color[rgb]{ 0,  .69,  .314}$\uparrow$ \textcolor[rgb]{ 0,  .69,  .314}{3.74} \\
          & rel.  & 12.09 & 12.08 & \color{red}$\downarrow$ \textcolor[rgb]{ 1,  0,  0}{0.11} & 22.36 & 22.31 & \color{red}$\downarrow$ \textcolor[rgb]{ 1,  0,  0}{0.22} & 6.93  & 6.92  & \color{red}$\downarrow$ \textcolor[rgb]{ 1,  0,  0}{0.03} \\
    \midrule
    \multirow{2}[2]{*}{\typetwo} & abs.  & 11.06 & 10.48 & \color{red}$\downarrow$ \textcolor[rgb]{ 1,  0,  0}{5.31} & 18.74 & 17.27 & \color{red}$\downarrow$ \textcolor[rgb]{ 1,  0,  0}{7.85} & 5.59  & 4.98  & \color{red}$\downarrow$ \textcolor[rgb]{ 1,  0,  0}{10.84} \\
          & rel.  & 11.04 & 10.49 & \color{red}$\downarrow$ \textcolor[rgb]{ 1,  0,  0}{5.03} & 19.18 & 17.56 & \color{red}$\downarrow$ \textcolor[rgb]{ 1,  0,  0}{8.45} & 6.11  & 5.14  & \color{red}$\downarrow$ \textcolor[rgb]{ 1,  0,  0}{15.86} \\
    \midrule
    \multirow{2}[2]{*}{\typethree} & abs.  & 12.73 & 12.68 & \color{red}$\downarrow$ \textcolor[rgb]{ 1,  0,  0}{0.40} & 23.30 & 23.19 & \color{red}$\downarrow$ \textcolor[rgb]{ 1,  0,  0}{0.45} & 6.93  & 7.02  & \color[rgb]{ 0,  .69,  .314}$\uparrow$ \textcolor[rgb]{ 0,  .69,  .314}{1.35} \\
          & rel.  & 12.93 & 12.93 & \color{red}$\downarrow$ \textcolor[rgb]{ 1,  0,  0}{0.03} & 23.55 & 23.50 & \color{red}$\downarrow$ \textcolor[rgb]{ 1,  0,  0}{0.21} & 7.53  & 7.58  & \color[rgb]{ 0,  .69,  .314}$\uparrow$ \textcolor[rgb]{ 0,  .69,  .314}{0.72} \\
    \midrule
    \multirow{2}[2]{*}{\typefour} & abs.  & 12.92 & 12.83 & \color{red}$\downarrow$ \textcolor[rgb]{ 1,  0,  0}{0.63} & 22.94 & 22.82 & \color{red}$\downarrow$ \textcolor[rgb]{ 1,  0,  0}{0.50} & 6.78  & 6.86  & \color[rgb]{ 0,  .69,  .314}$\uparrow$ \textcolor[rgb]{ 0,  .69,  .314}{1.17} \\
          & rel.  & 13.18 & 13.16 & \color{red}$\downarrow$ \textcolor[rgb]{ 1,  0,  0}{0.16} & 23.75 & 23.68 & \color{red}$\downarrow$ \textcolor[rgb]{ 1,  0,  0}{0.31} & 7.63  & 7.57  & \color{red}$\downarrow$ \textcolor[rgb]{ 1,  0,  0}{0.78} \\
    \midrule
    \multirow{2}[2]{*}{\typefive} & abs.  & 13.80 & 12.69 & \color{red}$\downarrow$ \textcolor[rgb]{ 1,  0,  0}{8.02} & 24.44 & 21.77 & \color{red}$\downarrow$ \textcolor[rgb]{ 1,  0,  0}{10.95} & 7.53  & 5.04  & \color{red}$\downarrow$ \textcolor[rgb]{ 1,  0,  0}{33.10} \\
          & rel.  & 13.70 & 12.86 & \color{red}$\downarrow$ \textcolor[rgb]{ 1,  0,  0}{6.13} & 24.93 & 22.48 & \color{red}$\downarrow$ \textcolor[rgb]{ 1,  0,  0}{9.84} & 8.29  & 6.42  & \color{red}$\downarrow$ \textcolor[rgb]{ 1,  0,  0}{22.61} \\
    \midrule
    \multirow{2}[2]{*}{\typeall} & abs.  & 12.45 & 12.10 & \color{red}$\downarrow$ \textcolor[rgb]{ 1,  0,  0}{2.78} & 22.25 & 21.40 & \color{red}$\downarrow$ \textcolor[rgb]{ 1,  0,  0}{3.83} & 6.58  & 6.04  & \color{red}$\downarrow$ \textcolor[rgb]{ 1,  0,  0}{8.20} \\
          & rel.  & 12.59 & 12.30 & \color{red}$\downarrow$ \textcolor[rgb]{ 1,  0,  0}{2.28} & 22.75 & 21.91 & \color{red}$\downarrow$ \textcolor[rgb]{ 1,  0,  0}{3.73} & 7.30  & 6.72  & \color{red}$\downarrow$ \textcolor[rgb]{ 1,  0,  0}{7.82} \\
    \midrule
    \midrule
    \multicolumn{2}{c|}{Python} & \multicolumn{3}{c|}{BLEU} & \multicolumn{3}{c|}{ROUGE-L} & \multicolumn{3}{c}{METEOR} \\
    Type  & Pos.  & w/o. t. & w. t  & Imp. (\%)  & w/o. t. & w. t  & Imp.  & w/o. t. & w. t  & Imp. (\%) \\
    \midrule
    \multirow{2}[2]{*}{\typeone} & abs.  & 12.83 & 12.72 & \color{red}$\downarrow$ \textcolor[rgb]{ 1,  0,  0}{0.90} & 21.12 & 21.00 & \color{red}$\downarrow$ \textcolor[rgb]{ 1,  0,  0}{0.53} & 5.60  & 5.49  & \color{red}$\downarrow$ \textcolor[rgb]{ 1,  0,  0}{1.87} \\
          & rel.  & 13.49 & 13.37 & \color{red}$\downarrow$ \textcolor[rgb]{ 1,  0,  0}{0.93} & 22.46 & 22.26 & \color{red}$\downarrow$ \textcolor[rgb]{ 1,  0,  0}{0.88} & 6.76  & 6.58  & \color{red}$\downarrow$ \textcolor[rgb]{ 1,  0,  0}{2.73} \\
    \midrule
    \multirow{2}[2]{*}{\typetwo} & abs.  & 11.43 & 10.83 & \color{red}$\downarrow$ \textcolor[rgb]{ 1,  0,  0}{5.21} & 18.34 & 16.70 & \color{red}$\downarrow$ \textcolor[rgb]{ 1,  0,  0}{8.95} & 5.06  & 4.36  & \color{red}$\downarrow$ \textcolor[rgb]{ 1,  0,  0}{13.89} \\
          & rel.  & 11.79 & 11.34 & \color{red}$\downarrow$ \textcolor[rgb]{ 1,  0,  0}{3.81} & 19.02 & 17.71 & \color{red}$\downarrow$ \textcolor[rgb]{ 1,  0,  0}{6.87} & 5.86  & 5.10  & \color{red}$\downarrow$ \textcolor[rgb]{ 1,  0,  0}{13.09} \\
    \midrule
    \multirow{2}[2]{*}{\typethree} & abs.  & 12.83 & 12.78 & \color{red}$\downarrow$ \textcolor[rgb]{ 1,  0,  0}{0.41} & 21.60 & 21.53 & \color{red}$\downarrow$ \textcolor[rgb]{ 1,  0,  0}{0.29} & 5.59  & 5.51  & \color{red}$\downarrow$ \textcolor[rgb]{ 1,  0,  0}{1.34} \\
          & rel.  & 13.45 & 13.40 & \color{red}$\downarrow$ \textcolor[rgb]{ 1,  0,  0}{0.40} & 22.90 & 22.79 & \color{red}$\downarrow$ \textcolor[rgb]{ 1,  0,  0}{0.49} & 7.08  & 6.95  & \color{red}$\downarrow$ \textcolor[rgb]{ 1,  0,  0}{1.77} \\
    \midrule
    \multirow{2}[2]{*}{\typefour} & abs.  & 13.18 & 12.99 & \color{red}$\downarrow$ \textcolor[rgb]{ 1,  0,  0}{1.41} & 21.09 & 20.60 & \color{red}$\downarrow$ \textcolor[rgb]{ 1,  0,  0}{2.34} & 4.78  & 4.52  & \color{red}$\downarrow$ \textcolor[rgb]{ 1,  0,  0}{5.41} \\
          & rel.  & 13.71 & 13.58 & \color{red}$\downarrow$ \textcolor[rgb]{ 1,  0,  0}{0.94} & 22.40 & 22.07 & \color{red}$\downarrow$ \textcolor[rgb]{ 1,  0,  0}{1.50} & 6.42  & 6.29  & \color{red}$\downarrow$ \textcolor[rgb]{ 1,  0,  0}{1.91} \\
    \midrule
    \multirow{2}[2]{*}{\typefive} & abs.  & 13.84 & 13.17 & \color{red}$\downarrow$ \textcolor[rgb]{ 1,  0,  0}{4.83} & 23.06 & 21.41 & \color{red}$\downarrow$ \textcolor[rgb]{ 1,  0,  0}{7.13} & 6.52  & 5.37  & \color{red}$\downarrow$ \textcolor[rgb]{ 1,  0,  0}{17.71} \\
          & rel.  & 14.33 & 13.82 & \color{red}$\downarrow$ \textcolor[rgb]{ 1,  0,  0}{3.57} & 24.11 & 22.71 & \color{red}$\downarrow$ \textcolor[rgb]{ 1,  0,  0}{5.80} & 7.78  & 6.85  & \color{red}$\downarrow$ \textcolor[rgb]{ 1,  0,  0}{11.99} \\
    \midrule
    \multirow{2}[2]{*}{\typeall} & abs.  & 12.82 & 12.50 & \color{red}$\downarrow$ \textcolor[rgb]{ 1,  0,  0}{2.52} & 21.04 & 20.25 & \color{red}$\downarrow$ \textcolor[rgb]{ 1,  0,  0}{3.76} & 5.51  & 5.05  & \color{red}$\downarrow$ \textcolor[rgb]{ 1,  0,  0}{8.34} \\
          & rel.  & 13.36 & 13.10 & \color{red}$\downarrow$ \textcolor[rgb]{ 1,  0,  0}{1.90} & 22.18 & 21.51 & \color{red}$\downarrow$ \textcolor[rgb]{ 1,  0,  0}{3.02} & 6.78  & 6.35  & \color{red}$\downarrow$ \textcolor[rgb]{ 1,  0,  0}{6.29} \\
    \bottomrule
    \end{tabular}%
    }
  \label{tab:csm_text}%
\end{table*}%

\textbf{\alinemdf{} (\typethree).} 
We find that seq-based Transformer shows robust performance under
\textit{\linemdf}.
For instance, the impact of this type of code transformation on the code search task is 0.30\% and -0.87\% for Java and Python (seen in Table~\ref{tab:cs_text}), respectively.

\textbf{\atokensys{} (\typefour).} 
From the experimental results, we observe that the \textit{\tokensys} has a slight
influence on all tasks.
For example, the MRR score has a decrease of 1.02\% and 3.19\% respectively for Java and Python on the code completion task, respectively (seen in Table~\ref{tab:cc_text}).

\textbf{\atokenuser{} (\typefive).} 
We observe that this type of code transformation has a substantial impact on all code intelligence tasks. For instance, the MRR score has a decrease of 42.21\% and 38.22\% for Java and Python under \textit{\tokenuser{}} in code search task, respectively (seen in Table~\ref{tab:cs_text}).

\begin{tcolorbox}[breakable,width=\linewidth,boxrule=0pt,top=1pt, bottom=1pt, left=1pt,right=1pt, colback=black!15,colframe=gray!20]
\textbf{Finding 1:} 
Transformer's performance is affected to varying degrees by different types of code transformations. 

\end{tcolorbox}

Based on the above analysis, we achieve that seq-based Transformer shows robust performance under \textit{\block{}}, \textit{\linemdf{}} and \textit{\tokensys{}}, but suffers from obvious performance degradation under \textit{\lineadd{}} and \textit{\tokenuser{}}. For example, 
in code completion task, 
the decreases of MRR score for Java under \textit{\lineadd{}} and \textit{\tokenuser{}} are 10.62\% and 9.54\%, respectively, while the decreases caused by \textit{\block{}}, \textit{\linemdf{}} and \textit{\tokensys{}} are 0.81\%, 1.66\%, and 1.02\%, respectively.

\begin{tcolorbox}[breakable,width=\linewidth,boxrule=0pt,top=1pt, bottom=1pt, left=1pt,right=1pt, colback=black!15,colframe=gray!20]
\textbf{Finding 2:} The \lineadd{} and \tokenuser{} present greatest impact on the performance of seq-based Transformer.
\end{tcolorbox}

During experimentation, we also find that inserting junk code at different locations (\textit{ID-2 front / middle / end}) has different impact on the performance of seq-based Transformer. 
\textit{ID-2 middle} represents the cases where the junk code is inserted inside the given code, and it is also the default implementation of \textit{ID-2} in other code transformation experiments.
\textit{ID-2 front / end} represents the junk code inserted before and after the given code, respectively.
The results
of inserting junk code at different locations for the code search and summarization tasks are shown in Table~\ref{tab:junkcode}.
We observe that Transformer's performance degrades more evidently when junk code is inserted at the front of the given code than when it is inserted randomly or at the end.
For example, on the code search task, the decrease of the ID-2 front / middle / end on seq-based Transformer are 42.67\% / 25.31\% / 15.95\% for Java respectively, while the decrease for Python are 88.19\% / 14.78\% / 10.56\%, respectively.

\begin{table}[htbp]
  \centering
  \caption{Impact of different insert locations of junk code on seq-based Transformer. CS and CSM represent code search and code summarization tasks, respectively. The ``Loc.'' represents the insert locations, and middle / front / end represent inserting junk code to the given code at the middle / at the front / at the end, respectively. The \textcolor{red}{red} color indicates the degree of decrease.}
  \begin{adjustbox}{width=.49\textwidth}
    \begin{tabular}{cc|ccc|ccc}
    \toprule
    \multicolumn{2}{c|}{CS} & \multicolumn{3}{c|}{MRR (Java)} & \multicolumn{3}{c}{MRR (Python)} \\
    Loc.  & Pos.  & w/o. t. & w. t. & imp. (\%) & w/o. t. & w. t. & imp. (\%) \\
    \midrule
        \multirow{2}[2]{*}{front} & abs.  & 0.4404 & 0.2525 & \color{red}$\downarrow$ \textcolor[rgb]{ 1,  0,  0}{42.67} & 0.3901 & 0.0461 & \color{red}$\downarrow$ \textcolor[rgb]{ 1,  0,  0}{88.19} \\
          & rel.  & 0.4461 & 0.2606 & \color{red}$\downarrow$ \textcolor[rgb]{ 1,  0,  0}{41.57} & 0.4117 & 0.0564 & \color{red}$\downarrow$ \textcolor[rgb]{ 1,  0,  0}{86.30} \\
    \midrule
    \multirow{2}[2]{*}{middle} & abs.  & 0.4404 & 0.3290 & \color{red}$\downarrow$ \textcolor[rgb]{ 1,  0,  0}{25.31} & 0.3901 & 0.3325 & \color{red}$\downarrow$ \textcolor[rgb]{ 1,  0,  0}{14.78} \\
          & rel.  & 0.4461 & 0.3305 & \color{red}$\downarrow$ \textcolor[rgb]{ 1,  0,  0}{25.91} & 0.4117 & 0.3324 & \color{red}$\downarrow$ \textcolor[rgb]{ 1,  0,  0}{19.26} \\
    \midrule

    \multirow{2}[2]{*}{end} & abs.  & 0.4404 & 0.3702 & \color{red}$\downarrow$ \textcolor[rgb]{ 1,  0,  0}{15.95} & 0.3901 & 0.3489 & \color{red}$\downarrow$ \textcolor[rgb]{ 1,  0,  0}{10.56} \\
          & rel.  & 0.4461 & 0.3748 & \color{red}$\downarrow$ \textcolor[rgb]{ 1,  0,  0}{15.98} & 0.4117 & 0.3456 & \color{red}$\downarrow$ \textcolor[rgb]{ 1,  0,  0}{16.05} \\
    \midrule
    \midrule
    \multicolumn{2}{c|}{CSM} & \multicolumn{3}{c|}{BLEU (Java)} & \multicolumn{3}{c}{BLEU (Python)} \\
    Loc.  & Pos.  & w/o. t. & w. t. & imp. (\%) & w/o. t. & w. t. & imp. (\%) \\
    \midrule
        \multirow{2}[2]{*}{front} & abs.  & 13.80 & 12.07 & \color{red}$\downarrow$ \textcolor[rgb]{ 1,  0,  0}{12.49} & 13.85 & 11.67 & \color{red}$\downarrow$ \textcolor[rgb]{ 1,  0,  0}{15.76} \\
          & rel.  & 13.70 & 12.29 & \color{red}$\downarrow$ \textcolor[rgb]{ 1,  0,  0}{10.31} & 14.35 & 12.43 & \color{red}$\downarrow$ \textcolor[rgb]{ 1,  0,  0}{13.38} \\
    \midrule
    \multirow{2}[2]{*}{middle} & abs.  & 13.80 & 12.13 & \color{red}$\downarrow$ \textcolor[rgb]{ 1,  0,  0}{12.06} & 13.85 & 12.27 & \color{red}$\downarrow$ \textcolor[rgb]{ 1,  0,  0}{11.41} \\
          & rel.  & 13.70 & 12.30 & \color{red}$\downarrow$ \textcolor[rgb]{ 1,  0,  0}{10.27} & 14.35 & 13.11 & \color{red}$\downarrow$ \textcolor[rgb]{ 1,  0,  0}{8.64} \\
    \midrule

    \multirow{2}[2]{*}{end} & abs.  & 13.80 & 12.24 & \color{red}$\downarrow$ \textcolor[rgb]{ 1,  0,  0}{11.28} & 13.85 & 12.47 & \color{red}$\downarrow$ \textcolor[rgb]{ 1,  0,  0}{9.99} \\
          & rel.  & 13.70 & 12.42 & \color{red}$\downarrow$ \textcolor[rgb]{ 1,  0,  0}{9.39} & 14.35 & 13.25 & \color{red}$\downarrow$ \textcolor[rgb]{ 1,  0,  0}{7.64} \\
    \bottomrule
    \end{tabular}%
    \end{adjustbox}
  \label{tab:junkcode}%
\end{table}%

\begin{tcolorbox}[breakable,width=\linewidth,boxrule=0pt,top=1pt, bottom=1pt, left=1pt,right=1pt, colback=black!15,colframe=gray!20]
\textbf{Finding 3:} Inserting junk code
at the front of the given code can affect Transformer's performance more
than inserting junk code at other locations.
\end{tcolorbox}

\subsubsection{Absolute position v.s. relative position} 

The section looks into whether position encoding
impacts
Transformer's performance under
code transformation.
From Table \ref{tab:cc_text}-\ref{tab:csm_text}, we find that Transformer with relative position encoding (\rltrans) shows more robust performance better compared to that with absolute position encoding (\abtrans) on three code intelligence tasks.
For example, the overall MRR score of \abtrans has a decrease of 4.50\% and 5.12\% for Java and Python on the code completion task, while the decrease of \rltrans is 3.67\% and 3.07\%, respectively. 
More specifically, we find that the influence of relative position encoding is more obvious under \textit{\lineadd} than other types of code transformation on the code completion task. 
For example, the \abtrans and \rltrans{}' MRR values under \textit{\lineadd} decrease by 10.62\% and 5.37\% for Java and 12.45\% and 3.65\% for Python, respectively, while the decreases under \textit{\linemdf} are are 1.66\% and 1.61\% for Java and 1.72\% v.s. 3.65\% for Python, respectively (\typetwo{} and \typethree{} in Table~\ref{tab:cc_text}).
The robustness improvement may be attributed to that the relative position encoding
enhances the relationship between the tokens with the adjacent preceding tokens\cite{shaw2018self}, making Transformer's predictions more accurate on code completion. 

\begin{tcolorbox}[breakable,width=\linewidth,boxrule=0pt,top=1pt, bottom=1pt, left=1pt,right=1pt, colback=black!15,colframe=gray!20]
\textbf{Finding 4:} Seq-based Transformer with relative position encoding shows more robust performance compared to that with absolute position encoding under most code transformations.
\end{tcolorbox}

To sum up, seq-based Transformer shows robust performance under \textit{\block{}}, \textit{\linemdf{}} and \textit{\tokensys{}}. However, \textit{\lineadd{}} and \textit{\tokenuser{}} have substantial impact on Transformer's performance. Besides, the relative position encoding can improve the robustness of seq-based Transformer under most code transformations.

\subsection{Answer to RQ2: Impact on AST-based Transformer}
\label{sec:q2}

In this section, we compare the performance of AST-based Transformer before and after different code transformations for three different tasks.
Table~\ref{tab:cc_ast}, Table~\ref{tab:cs_ast} and Table~\ref{tab:csm_ast} present the overall results of code completion, code search and code summarization on ASTs, respectively.

\begin{table}[htbp]
  \centering
  \caption{Results of code transformation on the performance of AST-based Transformer for the code completion task. The \textcolor{red}{red} and \textcolor[rgb]{ 0,  .69,  .314}{green} colors indicate the degree of decrease and increase, respectively.}
   \begin{adjustbox}{width=.49\textwidth}
    \begin{tabular}{cc|ccc|ccc}
    \toprule
    \multicolumn{2}{c|}{} & \multicolumn{3}{c|}{MRR (Java)} & \multicolumn{3}{c}{MRR (Python)} \\
    Type  & Pos.  & w/o. t. & w. t. & imp. (\%) & w/o. t. & w. t. & imp. (\%) \\
    \midrule
    \multirow{2}[2]{*}{\typeone} & abs.  & 0.8030 & 0.8043 & \color[rgb]{ 0,  .69,  .314}$\uparrow$ \textcolor[rgb]{ 0,  .69,  .314}{0.16} & 0.7630 & 0.7624 & \color{red}$\downarrow$ \textcolor[rgb]{ 1,  0,  0}{0.08} \\
          & rel.  & 0.8050 & 0.8066 & \color[rgb]{ 0,  .69,  .314}$\uparrow$ \textcolor[rgb]{ 0,  .69,  .314}{0.20} & 0.7646 & 0.7644 & \color{red}$\downarrow$ \textcolor[rgb]{ 1,  0,  0}{0.01} \\
    \midrule
    \multirow{2}[1]{*}{\typetwo} & abs.  & 0.6563 & 0.6346 & \color{red}$\downarrow$ \textcolor[rgb]{ 1,  0,  0}{3.31} & 0.6308 & 0.6208 & \color{red}$\downarrow$ \textcolor[rgb]{ 1,  0,  0}{1.59} \\
          & rel.  & 0.6578 & 0.6361 & \color{red}$\downarrow$ \textcolor[rgb]{ 1,  0,  0}{3.29} & 0.6325 & 0.6244 & \color{red}$\downarrow$ \textcolor[rgb]{ 1,  0,  0}{1.28} \\
    \midrule
    \multirow{2}[1]{*}{\typethree} & abs.  & 0.7996 & 0.7951 & \color{red}$\downarrow$ \textcolor[rgb]{ 1,  0,  0}{0.55} & 0.7554 & 0.7504 & \color{red}$\downarrow$ \textcolor[rgb]{ 1,  0,  0}{0.67} \\
          & rel.  & 0.8026 & 0.7983 & \color{red}$\downarrow$ \textcolor[rgb]{ 1,  0,  0}{0.53} & 0.7577 & 0.7526 & \color{red}$\downarrow$ \textcolor[rgb]{ 1,  0,  0}{0.67} \\
    \midrule
    \multirow{2}[1]{*}{\typefour} & abs.  & 0.8001 & 0.7941 & \color{red}$\downarrow$ \textcolor[rgb]{ 1,  0,  0}{0.75} & 0.7531 & 0.7444 & \color{red}$\downarrow$ \textcolor[rgb]{ 1,  0,  0}{1.17} \\
          & rel.  & 0.8017 & 0.7961 & \color{red}$\downarrow$ \textcolor[rgb]{ 1,  0,  0}{0.70} & 0.7547 & 0.7477 & \color{red}$\downarrow$ \textcolor[rgb]{ 1,  0,  0}{0.92} \\
  \midrule
    \multirow{2}[1]{*}{\typefive} & abs.  & 0.7896 & 0.7201 & \color{red}$\downarrow$ \textcolor[rgb]{ 1,  0,  0}{8.81} & 0.7569 & 0.7098 & \color{red}$\downarrow$ \textcolor[rgb]{ 1,  0,  0}{6.23} \\
          & rel.  & 0.7916 & 0.7210 & \color{red}$\downarrow$ \textcolor[rgb]{ 1,  0,  0}{8.91} & 0.7589 & 0.7107 & \color{red}$\downarrow$ \textcolor[rgb]{ 1,  0,  0}{6.35} \\
    \midrule
    \multirow{2}[2]{*}{\typeall} & abs.  & 0.7697 & 0.7496 & \color{red}$\downarrow$ \textcolor[rgb]{ 1,  0,  0}{2.61} & 0.7319 & 0.7175 & \color{red}$\downarrow$ \textcolor[rgb]{ 1,  0,  0}{1.95} \\
          & rel.  & 0.7717 & 0.7516 & \color{red}$\downarrow$ \textcolor[rgb]{ 1,  0,  0}{2.60} & 0.7337 & 0.7200 & \color{red}$\downarrow$ \textcolor[rgb]{ 1,  0,  0}{1.87} \\
    \bottomrule
    \end{tabular}%
    \end{adjustbox}
  \label{tab:cc_ast}%
\end{table}%

\begin{table}[htbp]
  \centering
  \caption{Results of code transformation on the performance of AST-based Transformer for the code search task. The \textcolor{red}{red} and \textcolor[rgb]{ 0,  .69,  .314}{green} colors indicate the degree of decrease and increase, respectively.}
     \begin{adjustbox}{width=.49\textwidth}

    \begin{tabular}{cc|ccc|ccc}
    \toprule
    \multicolumn{2}{c|}{} & \multicolumn{3}{c|}{MRR (Java)} & \multicolumn{3}{c}{MRR (Python)} \\
    Type  & Pos.  & w/o. t. & w. t. & imp. (\%) & w/o. t. & w. t. & imp. (\%) \\
    \midrule
    \multirow{2}[2]{*}{\typeone} & abs.  & 0.3594 & 0.3599 & \color[rgb]{ 0,  .69,  .314}$\uparrow$ \textcolor[rgb]{ 0,  .69,  .314}{0.14} & 0.3990 & 0.3997 & \color[rgb]{ 0,  .69,  .314}$\uparrow$ \textcolor[rgb]{ 0,  .69,  .314}{0.18} \\
          & rel.  & 0.3698 & 0.3702 & \color[rgb]{ 0,  .69,  .314}$\uparrow$ \textcolor[rgb]{ 0,  .69,  .314}{0.09} & 0.3957 & 0.3953 & \color{red}$\downarrow$ \textcolor[rgb]{ 1,  0,  0}{0.10} \\
    \midrule
    \multirow{2}[1]{*}{\typetwo} & abs.  & 0.3368 & 0.3220 & \color{red}$\downarrow$ \textcolor[rgb]{ 1,  0,  0}{4.41} & 0.2898 & 0.2238 & \color{red}$\downarrow$ \textcolor[rgb]{ 1,  0,  0}{22.75} \\
          & rel.  & 0.3625 & 0.3462 & \color{red}$\downarrow$ \textcolor[rgb]{ 1,  0,  0}{4.49} & 0.2847 & 0.2168 & \color{red}$\downarrow$ \textcolor[rgb]{ 1,  0,  0}{23.85} \\
          \midrule
    \multirow{2}[1]{*}{\typethree} & abs.  & 0.3978 & 0.3980 & \color[rgb]{ 0,  .69,  .314}$\uparrow$ \textcolor[rgb]{ 0,  .69,  .314}{0.04} & 0.4111 & 0.4097 & \color{red}$\downarrow$ \textcolor[rgb]{ 1,  0,  0}{0.35} \\
          & rel.  & 0.4225 & 0.4220 & \color{red}$\downarrow$ \textcolor[rgb]{ 1,  0,  0}{0.13} & 0.4057 & 0.4046 & \color{red}$\downarrow$ \textcolor[rgb]{ 1,  0,  0}{0.29} \\
    \midrule
    \multirow{2}[1]{*}{\typefour} & abs.  & 0.4388 & 0.4366 & \color{red}$\downarrow$ \textcolor[rgb]{ 1,  0,  0}{0.49} & 0.3924 & 0.3448 & \color{red}$\downarrow$ \textcolor[rgb]{ 1,  0,  0}{12.14} \\
          & rel.  & 0.4617 & 0.4607 & \color{red}$\downarrow$ \textcolor[rgb]{ 1,  0,  0}{0.21} & 0.3910 & 0.3415 & \color{red}$\downarrow$ \textcolor[rgb]{ 1,  0,  0}{12.67} \\
          \midrule
    \multirow{2}[1]{*}{\typefive} & abs.  & 0.4287 & 0.2428 & \color{red}$\downarrow$ \textcolor[rgb]{ 1,  0,  0}{43.36} & 0.4245 & 0.2644 & \color{red}$\downarrow$ \textcolor[rgb]{ 1,  0,  0}{37.71} \\
          & rel.  & 0.4439 & 0.2538 & \color{red}$\downarrow$ \textcolor[rgb]{ 1,  0,  0}{42.83} & 0.4235 & 0.2678 & \color{red}$\downarrow$ \textcolor[rgb]{ 1,  0,  0}{36.77} \\
    \midrule
    \multirow{2}[2]{*}{\typeall} & abs.  & 0.3923 & 0.3519 & \color{red}$\downarrow$ \textcolor[rgb]{ 1,  0,  0}{10.31} & 0.3834 & 0.3285 & \color{red}$\downarrow$ \textcolor[rgb]{ 1,  0,  0}{14.32} \\
          & rel.  & 0.4121 & 0.3706 & \color{red}$\downarrow$ \textcolor[rgb]{ 1,  0,  0}{10.08} & 0.3801 & 0.3252 & \color{red}$\downarrow$ \textcolor[rgb]{ 1,  0,  0}{14.46} \\
    \bottomrule
    \end{tabular}%
    \end{adjustbox}
  \label{tab:cs_ast}%
\end{table}%

\begin{table*}[htbp]
  \centering
  \caption{Results of code transformation on the performance of AST-based Transformer for the code summarization task. The above part presents the results of Transformer with absolute/relative position encoding strategies (``abs.'' and ``rel.'') for Java, and the below part presents results for Python. The \textcolor{red}{red} and \textcolor[rgb]{ 0,  .69,  .314}{green} colors indicate the degree of decrease and increase, respectively.}
  \scalebox{0.9}{
    \begin{tabular}{cc|ccc|ccc|ccc}
    \toprule
    \multicolumn{2}{c|}{Java} & \multicolumn{3}{c|}{BLEU} & \multicolumn{3}{c|}{ROUGE-L} & \multicolumn{3}{c}{METEOR} \\
    Type  & Pos.  & w/o. t. & w. t  & Imp. (\%) & w/o. t. & w. t  & Imp. (\%) & w/o. t. & w. t  & Imp. (\%) \\
    \midrule
    \multirow{2}[2]{*}{\typeone} & abs.  & 11.89 & 11.89 & $\rightarrow$ 0.00 & 21.85 & 21.81 & \color{red}$\downarrow$ \textcolor[rgb]{ 1,  0,  0}{0.19} & 6.91  & 7.01  & \color[rgb]{ 0,  .69,  .314}$\uparrow$ \textcolor[rgb]{ 0,  .69,  .314}{1.39} \\
          & rel.  & 12.40 & 12.43 & \color[rgb]{ 0,  .69,  .314}$\uparrow$ \textcolor[rgb]{ 0,  .69,  .314}{0.22} & 22.92 & 22.92 & \color{red}$\downarrow$ \textcolor[rgb]{ 1,  0,  0}{0.01} & 7.53  & 7.54  & \color[rgb]{ 0,  .69,  .314}$\uparrow$ \textcolor[rgb]{ 0,  .69,  .314}{0.09} \\
    \midrule
    \multirow{2}[2]{*}{\typetwo} & abs.  & 11.03 & 10.31 & \color{red}$\downarrow$ \textcolor[rgb]{ 1,  0,  0}{6.45} & 18.59 & 16.61 & \color{red}$\downarrow$ \textcolor[rgb]{ 1,  0,  0}{10.62} & 5.86  & 4.87  & \color{red}$\downarrow$ \textcolor[rgb]{ 1,  0,  0}{16.88} \\
          & rel.  & 11.10 & 10.51 & \color{red}$\downarrow$ \textcolor[rgb]{ 1,  0,  0}{5.31} & 19.10 & 16.99 & \color{red}$\downarrow$ \textcolor[rgb]{ 1,  0,  0}{11.02} & 6.17  & 5.20  & \color{red}$\downarrow$ \textcolor[rgb]{ 1,  0,  0}{15.71} \\
    \midrule
    \multirow{2}[2]{*}{\typethree} & abs.  & 12.70 & 12.84 & \color[rgb]{ 0,  .69,  .314}$\uparrow$ \textcolor[rgb]{ 0,  .69,  .314}{1.14} & 23.06 & 23.32 & \color[rgb]{ 0,  .69,  .314}$\uparrow$ \textcolor[rgb]{ 0,  .69,  .314}{1.12} & 7.67  & 7.76  & \color[rgb]{ 0,  .69,  .314}$\uparrow$ \textcolor[rgb]{ 0,  .69,  .314}{1.18} \\
          & rel.  & 12.99 & 13.00 & \color[rgb]{ 0,  .69,  .314}$\uparrow$ \textcolor[rgb]{ 0,  .69,  .314}{0.12} & 23.40 & 23.31 & \color{red}$\downarrow$ \textcolor[rgb]{ 1,  0,  0}{0.39} & 8.14  & 8.18  & \color[rgb]{ 0,  .69,  .314}$\uparrow$ \textcolor[rgb]{ 0,  .69,  .314}{0.43} \\
    \midrule
    \multirow{2}[2]{*}{\typefour} & abs.  & 12.93 & 12.88 & \color{red}$\downarrow$ \textcolor[rgb]{ 1,  0,  0}{0.39} & 23.03 & 22.76 & \color{red}$\downarrow$ \textcolor[rgb]{ 1,  0,  0}{1.17} & 7.38  & 7.34  & \color{red}$\downarrow$ \textcolor[rgb]{ 1,  0,  0}{0.52} \\
          & rel.  & 13.25 & 13.26 & \color[rgb]{ 0,  .69,  .314}$\uparrow$ \textcolor[rgb]{ 0,  .69,  .314}{0.08} & 23.65 & 23.52 & \color{red}$\downarrow$ \textcolor[rgb]{ 1,  0,  0}{0.55} & 8.06  & 7.93  & \color{red}$\downarrow$ \textcolor[rgb]{ 1,  0,  0}{1.56} \\
    \midrule
    \multirow{2}[2]{*}{\typefive} & abs.  & 13.94 & 12.92 & \color{red}$\downarrow$ \textcolor[rgb]{ 1,  0,  0}{7.31} & 24.65 & 22.05 & \color{red}$\downarrow$ \textcolor[rgb]{ 1,  0,  0}{10.53} & 8.24  & 6.77  & \color{red}$\downarrow$ \textcolor[rgb]{ 1,  0,  0}{17.93} \\
          & rel.  & 13.82 & 13.06 & \color{red}$\downarrow$ \textcolor[rgb]{ 1,  0,  0}{5.53} & 24.99 & 22.90 & \color{red}$\downarrow$ \textcolor[rgb]{ 1,  0,  0}{8.36} & 8.66  & 7.47  & \color{red}$\downarrow$ \textcolor[rgb]{ 1,  0,  0}{13.83} \\
    \midrule
    \multirow{2}[1]{*}{\typeall} & abs.  & 12.50 & 12.17 & \color{red}$\downarrow$ \textcolor[rgb]{ 1,  0,  0}{2.62} & 22.23 & 21.31 & \color{red}$\downarrow$ \textcolor[rgb]{ 1,  0,  0}{4.16} & 7.21  & 6.75  & \color{red}$\downarrow$ \textcolor[rgb]{ 1,  0,  0}{6.43} \\
          & rel.  & 12.71 & 12.45 & \color{red}$\downarrow$ \textcolor[rgb]{ 1,  0,  0}{2.05} & 22.81 & 21.93 & \color{red}$\downarrow$ \textcolor[rgb]{ 1,  0,  0}{3.87} & 7.71  & 7.26  & \color{red}$\downarrow$ \textcolor[rgb]{ 1,  0,  0}{5.84} \\
    \midrule
    \midrule 
    \multicolumn{2}{c|}{Python} & \multicolumn{3}{c|}{BLEU} & \multicolumn{3}{c|}{ROUGE-L} & \multicolumn{3}{c}{METEOR} \\
    Type  & Pos.  & w/o. t. & w. t  & Imp. (\%) & w/o. t. & w. t  & Imp. (\%) & w/o. t. & w. t  & Imp. (\%) \\
    \midrule
    \multirow{2}[2]{*}{\typeone} & abs.  & 12.51 & 12.50 & \color{red}$\downarrow$ \textcolor[rgb]{ 1,  0,  0}{0.06} & 18.89 & 18.88 & \color{red}$\downarrow$ \textcolor[rgb]{ 1,  0,  0}{0.09} & 4.85  & 4.83  & \color{red}$\downarrow$ \textcolor[rgb]{ 1,  0,  0}{0.44} \\
          & rel.  & 13.11 & 13.05 & \color{red}$\downarrow$ \textcolor[rgb]{ 1,  0,  0}{0.47} & 19.66 & 19.60 & \color{red}$\downarrow$ \textcolor[rgb]{ 1,  0,  0}{0.29} & 5.94  & 5.94  & \color[rgb]{ 0,  .69,  .314}$\uparrow$ \textcolor[rgb]{ 0,  .69,  .314}{0.10} \\
    \midrule
    \multirow{2}[2]{*}{\typetwo} & abs.  & 10.72 & 10.59 & \color{red}$\downarrow$ \textcolor[rgb]{ 1,  0,  0}{1.15} & 15.90 & 15.54 & \color{red}$\downarrow$ \textcolor[rgb]{ 1,  0,  0}{2.25} & 4.16  & 3.98  & \color{red}$\downarrow$ \textcolor[rgb]{ 1,  0,  0}{4.21} \\
          & rel.  & 11.47 & 11.13 & \color{red}$\downarrow$ \textcolor[rgb]{ 1,  0,  0}{2.99} & 16.70 & 15.78 & \color{red}$\downarrow$ \textcolor[rgb]{ 1,  0,  0}{5.50} & 5.11  & 4.51  & \color{red}$\downarrow$ \textcolor[rgb]{ 1,  0,  0}{11.64} \\
    \midrule
    \multirow{2}[2]{*}{\typethree} & abs.  & 12.39 & 12.37 & \color{red}$\downarrow$ \textcolor[rgb]{ 1,  0,  0}{0.16} & 19.09 & 19.03 & \color{red}$\downarrow$ \textcolor[rgb]{ 1,  0,  0}{0.34} & 4.97  & 4.90  & \color{red}$\downarrow$ \textcolor[rgb]{ 1,  0,  0}{1.46} \\
          & rel.  & 12.97 & 12.93 & \color{red}$\downarrow$ \textcolor[rgb]{ 1,  0,  0}{0.35} & 19.79 & 19.71 & \color{red}$\downarrow$ \textcolor[rgb]{ 1,  0,  0}{0.41} & 5.97  & 5.97  & \color{red}$\downarrow$ \textcolor[rgb]{ 1,  0,  0}{0.01} \\
    \midrule
    \multirow{2}[2]{*}{\typefour} & abs.  & 12.76 & 12.72 & \color{red}$\downarrow$ \textcolor[rgb]{ 1,  0,  0}{0.38} & 19.13 & 19.01 & \color{red}$\downarrow$ \textcolor[rgb]{ 1,  0,  0}{0.68} & 4.74  & 4.78  & \color[rgb]{ 0,  .69,  .314}$\uparrow$ \textcolor[rgb]{ 0,  .69,  .314}{0.99} \\
          & rel.  & 13.41 & 13.31 & \color{red}$\downarrow$ \textcolor[rgb]{ 1,  0,  0}{0.73} & 19.79 & 19.74 & \color{red}$\downarrow$ \textcolor[rgb]{ 1,  0,  0}{0.26} & 5.68  & 5.59  & \color{red}$\downarrow$ \textcolor[rgb]{ 1,  0,  0}{1.65} \\
    \midrule
    \multirow{2}[2]{*}{\typefive} & abs.  & 12.93 & 12.78 & \color{red}$\downarrow$ \textcolor[rgb]{ 1,  0,  0}{1.16} & 19.69 & 19.25 & \color{red}$\downarrow$ \textcolor[rgb]{ 1,  0,  0}{2.23} & 5.35  & 4.93  & \color{red}$\downarrow$ \textcolor[rgb]{ 1,  0,  0}{7.79} \\
          & rel.  & 13.80 & 13.45 & \color{red}$\downarrow$ \textcolor[rgb]{ 1,  0,  0}{2.56} & 21.06 & 20.13 & \color{red}$\downarrow$ \textcolor[rgb]{ 1,  0,  0}{4.39} & 6.37  & 5.61  & \color{red}$\downarrow$ \textcolor[rgb]{ 1,  0,  0}{11.98} \\
    \midrule
    \multirow{2}[2]{*}{\typeall} & abs.  & 12.26 & 12.19 & \color{red}$\downarrow$ \textcolor[rgb]{ 1,  0,  0}{0.57} & 18.54 & 18.34 & \color{red}$\downarrow$ \textcolor[rgb]{ 1,  0,  0}{1.09} & 4.81  & 4.68  & \color{red}$\downarrow$ \textcolor[rgb]{ 1,  0,  0}{2.65} \\
          & rel.  & 12.95 & 12.77 & \color{red}$\downarrow$ \textcolor[rgb]{ 1,  0,  0}{1.39} & 19.40 & 18.99 & \color{red}$\downarrow$ \textcolor[rgb]{ 1,  0,  0}{2.10} & 5.81  & 5.52  & \color{red}$\downarrow$ \textcolor[rgb]{ 1,  0,  0}{4.98} \\
    \bottomrule
    \end{tabular}%
    }
  \label{tab:csm_ast}%
\end{table*}%

\subsubsection{Different types of code transformation on ASTs}
\label{sec:q2_type}

In this section, we analyze the effects of different types of code transformation on the performance of AST-based Transformer.
We observe that \textit{\lineadd{}} and \textit{\tokenuser{}} also have substantial impact on AST-based Transformer. For example, the decrease of AST-based Transformer on predicting Java's code are 3.31\% and 8.81\% under \textit{\lineadd{}} and \textit{\tokenuser{}}, respectively, while the decrease of that under \textit{\linemdf{}} and \textit{\tokensys{}} are 0.55\% and 0.75\%, respectively.
We elaborate on the detailed impact of ASTs on different types of code transformation in the following.

\textbf{\ablock{} (\typeone)} has a minor impact on AST-based Transformer's performance as Table~\ref{tab:cc_ast}-\ref{tab:csm_ast} shows. And AST-based Transformer shows more robust performance than seq-based Transformer.
For example, when it comes to generating Python's code summary, the values of BLEU, ROUGE-L, and METEOR decrease by 0.06\%, 0.09\%, and 0.44\% on AST-based Transformer, while the decreases on seq-based Transformer are 0.90\%, 0.53\%, 1.87\%, respectively.

\textbf{\alineadd{} (\typetwo)}. 
We find that incorporating ASTs into Transformer can increase the model's robustness under \textit{\lineadd{}} on the code completion task, the Python's code summarization task and the Java's code search task.
For example, the MRR score has a decrease of 10.62\% and 12.45\% for Java and Python on seq-based Transformer (Seen in Table~\ref{tab:cc_text}), while the decrease is 3.31\% and 1.59\% on AST-based Transformer (Seen in Table~\ref{tab:cc_ast}), respectively.

\begin{table}[htbp]
  \centering
  \caption{Impact of different insert locations of junk code on AST-based Transformer. CS and CSM represent code search and code summarization tasks, respectively. The ``Loc.'' represents the insert locations, and middle / front / end represent inserting junk code to the given code at the middle / at the front / at the end, respectively. The \textcolor{red}{red} color indicates the degree of decrease.}
    \begin{adjustbox}{width=.49\textwidth}
    \begin{tabular}{cc|ccc|ccc}
    \toprule
    \multicolumn{2}{c|}{CS} & \multicolumn{3}{c|}{MRR (Java)} & \multicolumn{3}{c}{MRR (Python)} \\
    Loc.  & Pos.  & w/o. t. & w. t. & imp. (\%) & w/o. t. & w. t. & imp. (\%) \\
    \midrule
    \multirow{2}[2]{*}{front} & abs.  & 0.4184 & 0.3565 & \color{red}$\downarrow$ \textcolor[rgb]{ 1,  0,  0}{14.79} & 0.4364 & 0.0601 & \color{red}$\downarrow$ \textcolor[rgb]{ 1,  0,  0}{86.23} \\
          & rel.  & 0.4242 & 0.3448 & \color{red}$\downarrow$ \textcolor[rgb]{ 1,  0,  0}{18.72} & 0.4200 & 0.0389 & \color{red}$\downarrow$ \textcolor[rgb]{ 1,  0,  0}{90.74} \\
    \midrule
    \multirow{2}[2]{*}{middle} & abs.  & 0.4184 & 0.3821 & \color{red}$\downarrow$ \textcolor[rgb]{ 1,  0,  0}{8.66} & 0.4364 & 0.3750 & \color{red}$\downarrow$ \textcolor[rgb]{ 1,  0,  0}{14.06} \\
          & rel.  & 0.4242 & 0.3826 & \color{red}$\downarrow$ \textcolor[rgb]{ 1,  0,  0}{9.81} & 0.4200 & 0.3564 & \color{red}$\downarrow$ \textcolor[rgb]{ 1,  0,  0}{15.13} \\
    \midrule
    \multirow{2}[2]{*}{end} & abs.  & 0.4184 & 0.4009 & \color{red}$\downarrow$ \textcolor[rgb]{ 1,  0,  0}{4.17} & 0.4364 & 0.3981 & \color{red}$\downarrow$ \textcolor[rgb]{ 1,  0,  0}{8.78} \\
          & rel.  & 0.4242 & 0.4026 & \color{red}$\downarrow$ \textcolor[rgb]{ 1,  0,  0}{5.10} & 0.4200 & 0.3860 & \color{red}$\downarrow$ \textcolor[rgb]{ 1,  0,  0}{8.09} \\
    \midrule
    \midrule
    \multicolumn{2}{c|}{CSM} & \multicolumn{3}{c|}{BLEU (Java)} & \multicolumn{3}{c}{BLEU (Python)} \\
    Loc.  & Pos.  & w/o. t. & w. t. & imp. (\%) & w/o. t. & w. t. & imp. (\%) \\
    \midrule
        \multirow{2}[2]{*}{front} & abs.  & 13.94 & 12.66 & \color{red}$\downarrow$ \textcolor[rgb]{ 1,  0,  0}{9.18} & 12.93 & 12.47 & \color{red}$\downarrow$ \textcolor[rgb]{ 1,  0,  0}{3.56} \\
          & rel.  & 13.82 & 12.47 & \color{red}$\downarrow$ \textcolor[rgb]{ 1,  0,  0}{9.77} & 13.81 & 12.42 & \color{red}$\downarrow$ \textcolor[rgb]{ 1,  0,  0}{10.09} \\
    \midrule
    \multirow{2}[2]{*}{middle} & abs.  & 13.94 & 12.75 & \color{red}$\downarrow$ \textcolor[rgb]{ 1,  0,  0}{8.54} & 12.93 & 12.63 & \color{red}$\downarrow$ \textcolor[rgb]{ 1,  0,  0}{2.35} \\
          & rel.  & 13.82 & 12.68 & \color{red}$\downarrow$ \textcolor[rgb]{ 1,  0,  0}{8.25} & 13.81 & 13.08 & \color{red}$\downarrow$ \textcolor[rgb]{ 1,  0,  0}{5.28} \\
    \midrule

    \multirow{2}[2]{*}{end} & abs.  & 13.94 & 12.86 & \color{red}$\downarrow$ \textcolor[rgb]{ 1,  0,  0}{7.75} & 12.93 & 12.66 & \color{red}$\downarrow$ \textcolor[rgb]{ 1,  0,  0}{2.11} \\
          & rel.  & 13.82 & 12.80 & \color{red}$\downarrow$ \textcolor[rgb]{ 1,  0,  0}{7.40} & 13.81 & 13.12 & \color{red}$\downarrow$ \textcolor[rgb]{ 1,  0,  0}{5.02} \\
    \bottomrule
    \end{tabular}%
    \end{adjustbox}
  \label{tab:junkcode_ast}%
\end{table}%

In addition, inserting code snippets (\textit{ID-2}) at the front of the given code also affects AST-based Transformer's performance more than other locations as the seq-based Transformer. Besides, we observe that AST-based shows more robust performance under different locations of \textit{ID-2} than seq-based Transformer as shown in Table~\ref{tab:junkcode_ast}. For example, on the Java's code search task, the decrease of the \textit{ID-2 front / middle / end} on seq-based Transformer are 42.67\% / 25.31\% / 15.95\% respectively (Seen in Table~\ref{tab:junkcode}), while the decrease on AST-based Transformer are 14.79\% / 8.66\% / 4.17\%, respectively.

\textbf{\alinemdf{} (\typethree)} has a minor effect on AST-based Transformer. For instance, the MRR values decrease by 0.55\% and 0.67\%, respectively, when in Java's and Python's code completion.

\textbf{\atokensys{} (\typefour).} From the results, we also find that \textit{\tokensys} has a much smaller influence on AST-based Transformer than seq-based Transformer. For example, the MRR score of AST-based Transformer under \textit{\tokensys} has a decrease of 0.75\% and 1.17\% for Java and Python in code completion task, respectively, while the decrease of seq-based Transformer is 1.02\% and 3.19\%, respectively.

\textbf{\atokenuser{} (\typefive)} also results in a large quality drop of Transformer on traverse ASTs. We take the code search task under \textit{\tokenuser} as an example, AST-based Transformer have decreased MRR by 43.36\% and 37.71\% for Java and Python, respectively, while seq-based Transformer have decreased MRR by 42.21\% and 38.22\%. 
This results shows that AST-based Transformer is also vulnerable to \textit{\tokenuser} as seq-based Transformer.

\vspace{0.4cm}

Compared to the results of seq-based Transformer, we observe that utilizing ASTs can help Transformer maintain performance better.
For example, the overall MRR score of seq-based Transformer has a decrease of 4.50\% on the Java's code completion task (seen in Table~\ref{tab:cc_text}), while the decrease of AST-based Transformer is 2.61\%. Similar results can be seen in Python (5.12\% v.s. 1.95\%). In the code search task, the overall decreases of performance on seq-based Transformer are 13.73\% and 14.77\% (seen in Table~\ref{tab:cs_text}) for Java and Python, whereas the decreases on AST-based Transformer are 10.31\% and 14.32\% (seen in Table~\ref{tab:cs_ast}), respectively.

\begin{tcolorbox}[breakable,width=\linewidth,boxrule=0pt,top=1pt, bottom=1pt, left=1pt,right=1pt, colback=black!15,colframe=gray!20]
\textbf{Finding 5:} Compared to seq-based Transformer, AST-based Transformer demonstrates more robust performance under most code transformations.
\end{tcolorbox}

\subsubsection{Absolute position v.s. relative position on ASTs}

This section looks into whether position encoding has an impact on the AST-based Transformer's performance under code transformation.
From Table \ref{tab:cc_ast}-\ref{tab:csm_ast}, we find that relative position encoding does not make
obvious improvement on AST-based Transformer's robustness under code transformations. For example, the decrease in overall MRR score on Java's code completion task is 2.61\% and 2.60\% for absolute and relative position encoding, while the decrease on Python is 1.95\% and 1.87\%, respectively. Similar results can be seen on the code search task (10.31\% v.s. 10.08\% for Java and 14.32\% v.s. 14.46\% for Python).

\begin{tcolorbox}[breakable,width=\linewidth,boxrule=0pt,top=1pt, bottom=1pt, left=1pt,right=1pt, colback=black!15,colframe=gray!20]
\textbf{Finding 6:} Relative position encoding does not evidently improve the robustness of AST-based Transformer under code transformation.
\end{tcolorbox}

\vspace{0.4cm}

To sum up, the greatest impact of two types of code transformations on AST-based Transformer are also \lineadd{} and \tokenuser{}. 
And AST-based Transformer performs more robustly compared to seq-based Transformer.
Besides, relative position encoding does not evidently improve the robustness of AST-based Transformer when faced with code transformation. 
 
\section{DISCUSSION}
\label{sec:discuss}
The experiments in Section \ref{sec:result} have demonstrated that \textit{insertion/deletion transformation} and \textit{identifier transformation} have a substantial impact on the robustness of Transformer.
In this section, we discuss the Transformer's robustness from three perspectives: the input-ASTs, \textit{\lineadd}, and \textit{\tokenuser{}}.


\subsection{Effect
of ASTs on
the models' robustness}

Based on the experimental results in Section~\ref{sec:q2}, we observe that AST-based Transformer performs just marginally better than seq-based Transformer under different
code transformations. Besides, the relative position encoding does not obviously
improve the robustness of AST-based Transformer. Moreover, AST-based Transformers still cannot well understand the underlying semantic information of transformed code (e.g., \textit{\lineadd{} and \tokenuser}). The results indicate that exploiting ASTs for learning robust code semantics is still challenging.

We suggest that researchers can propose more effective
approaches to learn the code representations based on ASTs, and evaluate their impact on the robustness of code intelligence models.

\begin{tcolorbox}[breakable,width=\linewidth,boxrule=0.5pt,top=1pt, bottom=1pt, left=1pt,right=1pt, colback=white!15,colframe=black]
\textbf{Implication 1:} 
Transformer still faces challenges in exploiting ASTs for learning robust code semantics, and more exploration is needed to effectively capture
semantic information from ASTs.
\end{tcolorbox}

\subsection{Effect of \lineadd{}}

From the experimental results in Section~\ref{sec:q1} and \ref{sec:q2}, we observe that the \textit{\lineadd} has a substantial impact on all tasks. The results indicate that inserting some junk data (e.g., comments, code snippets, or libraries) can greatly bias Transformer-based code intelligence models. This may be because that the models fail to attend to the important part in the transformed
code.

One way to address the issue is to propose a more effective attention model for identifying the useful part in the input code. Another way is to
incorporate additional external knowledge such as API documentation to eliminate the distraction of \textit{\lineadd{}}.
For example, 
if Transformer-based models can recognize noisy part in the input based on the external information, the robustness of the models would be improved under the \textit{\lineadd}.

\begin{tcolorbox}[breakable,width=\linewidth,boxrule=0.5pt,top=1pt, bottom=1pt, left=1pt,right=1pt, colback=white!15,colframe=black]
\textbf{Implication 2:} 
Transformer needs a more effective attention approach or additional external knowledge to eliminate the distraction of noisy information
in the input code.
\end{tcolorbox}

\subsection{Effect of identifier transformation}

Since identifiers are an important part to express the naturalness of code, their importance
has been studied in many works~\cite{leclair2019neural, ahmed2018compilation}. Based on the experimental results in Section~\ref{sec:q1} and \ref{sec:q2}, we observe that the \textit{\tokenuser{}} have a substantial impact on both seq-based and AST-based Transformer. The results demonstrate the importance of identifiers for code intelligence tasks, and that it is hard for Transformer to learn the code semantics under \textit{identifier transformation}.

Identifiers could help models for better code understanding~\cite{efstathiou2019semantic,lacomis2019dire}.
However, most existing code intelligence models rely
on identifiers rather than underlying semantic information, resulting in models being vulnerable to \textit{identifier transformation}.
For example, some identifier-based adversarial attack strategies~\cite{zhang2020generating,yang2022natural} for code intelligence have been proposed in recent years. Future work is encouraged to explore how to effectively understand the code semantics under \textit{identifier transformation}.

\begin{tcolorbox}[breakable,width=\linewidth,boxrule=0.5pt,top=1pt, bottom=1pt, left=1pt,right=1pt, colback=white!15,colframe=black]
\textbf{Implication 3:} 
Future work is expected to well exploit the underlying semantics of identifiers rather than completely rely on the literal meanings of the identifiers.
\end{tcolorbox}


\section{THREATS TO VALIDITY}
\label{sec:therats}
In this section, we describe the possible threats we may face in this study and discuss how we mitigate them.

\textbf{Internal validity} is mainly about the data prepossessing and training models. 
To reduce the threats, we conduct experiments based
on the released code scripts,
and the default training hyper-parameters for all models. Besides, we run each experiment
for three
times and compute the average results for all tasks.

\textbf{Construct validity} is mainly about the suitability of our evaluation metrics. To reduce this risk, we select the most widely-used metrics for different tasks to evaluate the impact. For example, we use BLEU~\cite{papineni2002bleu}, ROUGE-L\cite{lin2004rouge} and Meteor\cite{banerjee2005meteor} to evaluate the impact of different transformations on code summarization task.

\textbf{External validity} is mainly concerned with dataset we use. In our experiments, we select Java and Python datasets from CodeSearchNet. However, CodeSearchNet has some problems that will affect our experiments. For example, some instances of code in CodeSearchNet cannot be parsed into an abstract syntax tree. To reduce the threats, we filter the datasets following the previous work\cite{lu2021codexglue}. To further reduce the threats, we plan to collect more open-source projects to reproduce our experiments.

\section{RELATED WORK}
\label{sec:related}

\subsection{Code intelligence task}

\textbf{Code completion task.} 
The success of deep learning boosts
a series of code completion works based on deep learning models\cite{svyatkovskiy2019pythia,wang2021code}.
For example, Alon et al. \cite{alon2020structural} proposed a structural language model based on Transformer,
which leverages the syntax to model the code snippet as a tree to complete code.
Liu et al. \cite{liu2020multi} pretrained a language model with a Transformer-based architecture and fine-tuned it for code completion.
Kim et al. \cite{kim2021code} proposed TravTrans, a Transformer-based approach that leverages ASTs for code completion.

\textbf{Code summarization task.} 
In recent years, many works that applied deep learning models in code summarization achieved great success and became more and more popular\cite{zhu2019automatic,wan2018improving}. For instance, Iyer et al.\cite{iyer2016summarizing} proposed an LSTM-based model with attention to generate code summaries for source code. Wei et al.\cite{wei2020retrieve} and Zhang et al.\cite{zhang2020retrieval} further introduced retrieved information for summarizing source code with the help of most similar code snippets. 
Alon et al. \cite{alon2018codeseq} sampled and encoded random AST paths into LSTMs to generate summaries of source code.
Ahmad et al.\cite{ahmad2020summarization} utilized Transformer on code summarization for better mapping the source code to their corresponding natural language summaries.

\textbf{Code search task.} The goal of code search is to find the most semantically related code from a collection of code based on a given natural language\cite{husain2019codesearchnet}. The traditional code search techniques are mainly based on information retrieval\cite{sindhgatta2006using,sim2011well}, while the most popular approaches in recent years are based on deep neural networks\cite{cambronero2019deep,gu2021multimodal}.
Sachdev et al.\cite{sachdev2018retrieval} proposed a neural code search engine which used a basic word–embedding for a code corpus.
Gu et al.\cite{gu2018deep} proposed an RNN-based code search model to represent source code and natural language queries through joint embedding. 
Fan et al.\cite{fang2021self} utilized a self-attention network to construct a code representation network for building the semantic relationship between code snippets and queries.
Besides, large-scale pre-train models such as Codebert\cite{feng2020codebert} and GraphCodeBERT\cite{guo2020graphcodebert} also demonstrated good performance on the code search task.

\subsection{Code transformation}

Code transformations are widely used for compiler optimizations\cite{cooper2006exploring,cavazos2007rapidly}, testability transformation\cite{binkley2011flagremover, mcminn2009empirical, baresel2004evolutionary}, refactoring\cite{daniel2007automated,mens2004survey}, etc. In recent years, there have been an amount of works in the field of code intelligence that employs code transformation.

\textbf{Data augmentation.} Yu et al. \cite{yu2022data} applied a source-to-source transformation strategy to increase the generalization capacity of deep learning models based on program-level data augmentation, and Wang et al.\cite{wang2021bridging} incorporated curriculum learning into program-level data augmentation in order to optimize the efficiency of fine-tuning pre-trained models. Bui et al.\cite{bui2021self} proposed a self-supervised contrastive learning framework to generate transformed code snippets and identify semantically-equivalent code snippets from large-scale unlabeled data of source code, which could generate additional dataset for fine-turning the pre-training models.

\textbf{Adversarial attack.} Quiring et al.\cite{quiring2019misleading} used semantics-preserving code transformations to generate adversarial examples and present a black-box attack that seeks examples by Monte-Carlo tree search. Li et al.\cite{li2021towards} leveraged code transformations to attack DL-based detectors and decouple feature learning and classifier learning to present a static analysis-based vulnerability detector. Zhang et al.\cite{zhang2020semantic} applied reinforcement learning to select semantics-preserving samples and proposed a black-box attack approach against GNN malware detection models

\textbf{Others applications.} Chen et al. \cite{chen2018learning} suggested a symbolic execution acceleration framework by machine-learning based compiler optimization tuning based on code transformation. Wu et al.\cite{wu2019automating} proposed an automating CUDA synchronization framework for bug detection at the LLVM-bitcode level via program transformation.

\subsection{Robustness of code intelligence}

Given the increased research interest in code intelligence, some works on the robustness of source code representation models are proposed\cite{yefet2020adversarial,rabin2021generalizability, bielik2020adversarial,ramakrishnan2020semantic,zhang2020generating,yang2022natural}. 

Yefet et al.\cite{yefet2020adversarial} proposed an adversarial attacks generation approach for code via gradient based optimization, and it was effective in generating both targeted and non-targeted attacks.
Rabin et al. \cite{rabin2021generalizability} defined the generalizability in neural program models and studied the generalizability of method name prediction models using six semantic-preserving changes.
Ramakrishnan et al. \cite{ramakrishnan2020semantic} proposed an adversarial-training method for neural models of code to enhance the robustness.
Zhang et al. \cite{zhang2020generating} defined the robust and non-robust features of DNNs, and proposed an identifier renaming algorithm (namely Metropolis-Hastings Modifier) for adversarial example generation on source code.
Yang et al. \cite{yang2022natural} designed a black-box attack approach that was aware of natural semantics when generating adversarial examples of code, which was better than that of Zhang et al. \cite{zhang2020generating}.
Bielik et al. \cite{bielik2020adversarial} refined the representation of source code and applied adversarial-training to improve robustness of neural models while preserving high accuracy.

\section{CONCLUSION AND FUTURE WORK}
\label{sec:conclusion}

In this study, we have empirically investigated the robustness and limitations of Transformer on code intelligence. 
We implement 27 and 24 code transformation strategies for Java and Python languages respectively and apply the transformed code to three code intelligence tasks to study the effect on Transformer.
Experimental results demonstrate that \lineadd{} and \tokenuser{} have the great impact on Transformer's performance.
Transformer based on ASTs shows more robust performance than the model based on only code sequence under most code transformations. Besides, the robustness of Transformer under code transformation is impacted by the design of positional encoding.
Based on the findings, we summarize some future directions for improving the robustness of Transformer on code intelligence. 

In the future, we plan to investigate the robustness of pre-trained models under code transformation.
Moreover, we plan to collect more open-source projects to reproduce our experiments and to support more programming languages in our code transformation.



\ifCLASSOPTIONcaptionsoff
  \newpage
\fi

\bibliographystyle{IEEEtran}
{\balance \tiny \bibliography{section_reference}}











\end{document}